\documentclass[manuscript]{emulateapj}
\usepackage{setspace}
\usepackage{graphicx}
\usepackage{natbib}
\usepackage{amsmath, amssymb}
\usepackage[colorlinks,urlcolor=blue,citecolor=blue,linkcolor=blue]{hyperref}

\shortauthors{Tamayo et al.}
\begin{document}

\title{Radial profiles of the Phoebe ring, a vast debris disk around Saturn}
\author{D. Tamayo\altaffilmark{1,2,3}, S. R. Markham\altaffilmark{4}, M. M. Hedman\altaffilmark{5}, J. A. Burns\altaffilmark{4} and D. P. Hamilton\altaffilmark{6}}
\altaffiltext{1}{Department of Physical \& Environmental Sciences, University of Toronto at Scarborough, Toronto, Ontario M1C 1A4, Canada}
\altaffiltext{2}{Canadian Institute for Theoretical Astrophysics, 60 St. George St, University of Toronto, Toronto, Ontario M5S 3H8, Canada}
\altaffiltext{3}{Centre for Planetary Sciences Fellow}
\altaffiltext{4}{Department of Astronomy, Cornell University, Ithaca, NY 14853, USA}
\altaffiltext{5}{Department of Physics, University of Idaho, Moscow, ID 83844, USA}
\altaffiltext{6}{Department of Astronomy, University of Maryland, College Park, MD 20742, USA}
\email{dtamayo@utoronto.ca}

\begin{abstract}
We present observations at optical wavelengths with the Cassini Spacecraft's Imaging Science System of the Phoebe ring, a vast debris disk around Saturn that seems to be collisionally generated by its irregular satellites.  The analysis reveals a radial profile from 80-260 Saturn radii ($R_S$) that changes behavior interior to $\approx 110 R_S$, which we attribute to either the moon Iapetus sweeping up small particles, or to orbital instabilities that cause the ring to flare up vertically. Our study yields an integrated I/F at 0.635 $\mu$m along Saturn's shadow in the Phoebe ring's midplane from 80-250 $R_S$ of $2.7^{+0.9}_{-0.3} \times 10^{-9}$.  We develop an analytical model for the size-dependent secular dynamics of retrograde Phoebe ring grains, and compare this model to the observations.  This analysis implies that 1) the ``Phoebe" ring is partially sourced by debris from irregular satellites beyond Phoebe's orbit and 2) the scattered light signal is dominated by small grains ($\lesssim 20\mu$m in size).  If we assume that the Phoebe ring is generated through steady-state micrometeoroid bombardment, this implies a power-law size distribution with index $> 4$, which is unusually steep among solar system rings. This suggests either a steep size distribution of ejecta when material is initially released, or a subsequent process that preferentially breaks up large grains.  
\end{abstract}


\section{Introduction}
Using the Spitzer infrared space telescope, \cite{Verbiscer09} discovered a vast dust ring around Saturn, far beyond the bright main rings.  
This debris disk was dubbed the Phoebe ring after the largest of Saturn's distant irregular satellites, which seems to be the dominant source for the material.
Approximately three dozen known irregular satellites \citep[see][for reviews]{Jewitt07, Nicholson08} form a swarm of mutually inclined, overlapping orbits---a relic of their capture process \citep{Pollack79, Nesvorny03, Cuk04, Cuk06, Nesvorny07}.
This led to a violent collisional history among these bodies continuing since early times \citep{Bottke10}.
Smaller collisions must be ongoing, both with circumplanetary objects too small to detect observationally, and with interplanetary meteoroids \citep[cf.,][]{Cuzzi98}.

While the disk is diffuse, the debris from these dark irregular satellites \citep{Grav15} can have important consequences.
Iapetus, the outermost of the large, tidally locked, regular satellites has a leading side approximately ten times darker than its trailing side.
Many years before its discovery, \cite{Soter74} (see also \citealt{Cruikshank83, Bell85, Buratti95}) hypothesized that inward transfer of such debris through Poynting-Robertson drag might explain Iapetus' stark hemispheric dichotomy.  
\cite{Burns96}, and more recently \cite{Tosi10} and \cite{Tamayo11}, showed that indeed, Iapetus should intercept most of the inspiraling material as it plows through the cloud, and that the longitudinal distribution of dark material on Iapetus can be well explained by dust infall under the action of radiation pressure.
Additionally, \cite{Spencer10, Denk10} showed that runaway ice sublimation and redeposition could accentuate initially subtle albedo differences to match the observed stark contrast.

Furthermore, this process of collisional grinding among the irregular satellites should be ubiquitous among the solar (and perhaps extrasolar) system's giant planets \citep{Bottke10, Kennedy11}, and this debris should also fall onto the respective outermost regular satellites.
Indeed, the Uranian regular satellites exhibit hemispherical color dichotomies \citep{Buratti91}, and \cite{Tamayo13a} showed that this could similarly be explained through dust infall, though the dynamics are additionally complicated by Uranus' extreme obliquity \citep{Tamayo13b}.  
\cite{Bottke13} argue the same process has occurred in the Jovian system.
As the only known debris disk sourced by irregular satellites, the Phoebe ring therefore presents a unique opportunity to learn about generic processes around giant planets, both in our solar system and beyond.

\cite{Tamayo14a}, hereafter THB14, detected the Phoebe ring's scattered light at optical wavelengths, using the Cassini spacecraft in orbit around Saturn.  
THB14 combined these optical measurements with the thermal emission data of \cite{Verbiscer09}, finding that Phoebe ring grains have low albedos similar to the dark irregular satellites \citep{Grav15}.
More recently, \cite{Hamilton15} combined detailed numerical models of dust grains' size-dependent spatial distributions with new data from the Wide-Field Infrared Survey Explorer (WISE) to extract the particle-size distribution in the disk.
They found that the Phoebe ring extends out to at least 270 Saturn radii\footnote{For this work we adopt $R_S$ = 60330 km, the convention used for calculating Saturn's gravitational moments.}($R_S$) and has a steep particle size distribution.
However, the Phoebe ring is so faint (normal optical depth $\sim 10^{-8}$) that scattered light from the planet dominates the signal inside $\approx 100$ Saturn radii ($R_S$).
This is too far out to detect an inner edge swept out by Iapetus, which orbits at $\approx 59 R_S$.

In this paper we present results from a new Cassini data set with a substantially higher signal-to-noise ratio than that of THB14.  
This renders the faint Phoebe ring signature clearly visible in our images, and we are able to additionally extract the Phoebe ring's radial structure.
We begin by presenting our data analysis, and by describing our data reduction methods in Sec.~\ref{methods}, and our results in Sec.~\ref{results}.
We then semi-analytically investigate the expected 3-D structure of the Phoebe ring, which should exhibit interesting dynamics closer to Iapetus, where the Sun stops being the dominant perturbation (as it is for grains at large Saturnocentric distances), and Saturn's oblateness becomes important. 
In Sec.~\ref{comparison} we compare our model to the data and we summarize our results in Sec.~\ref{conc}.

\section{Methods} \label{methods}
\subsection{Data Reduction}
The main observational challenge is that the scattered light signal from Phoebe ring grains is exceedingly weak (I/F $\sim 10^{-9}$).  Additionally, from Cassini's nearby vantage point, the Phoebe ring's thickness spans several tens of degrees; the Phoebe ring therefore appears as a uniform background across the 3.5'x3.5' field of view of Cassini's Imaging Science System (ISS) Wide-Angle Camera, WAC \citep{Porco04}.
We now briefly summarize the technique that THB14 developed to overcome these obstacles.  

The key is to detect the {\it deficit} of scattered light from unilluminated Phoebe ring grains lying in Saturn's shadow.  
Not only is the shadow narrow enough to be captured within a single WAC field of view, its apparent position relative to the background stars shifts as the spacecraft moves in its orbit.
THB14 examined several exposures of the same star field as Saturn's shadow moved through the images.
By subtracting images from one another, the constant background could be attenuated while retaining the moving shadow's signal.  

The signal-to-noise ratio can be substantially improved by positioning the spacecraft closer to the long axis of Saturn's shadow, which lengthens the column of Phoebe ring material along lines of sight that intersect the shadow (see Fig.\:1 in THB14).
On day 269 of 2013 (September 26$^{th}$), in Rev 197 (Cassini's 198$^{th}$ orbit about Saturn), we executed such an observation with Cassini only $\approx 6$ Saturn radii ($R_S$) from the shadow's axis (compared to $\approx 22 R_S$ in the observations of THB14).
We also maximized the shadow's movement across the field of view by taking images at the beginning and end of our observation window.

\begin{figure*}
\centering \resizebox{0.99\textwidth}{!}{\includegraphics{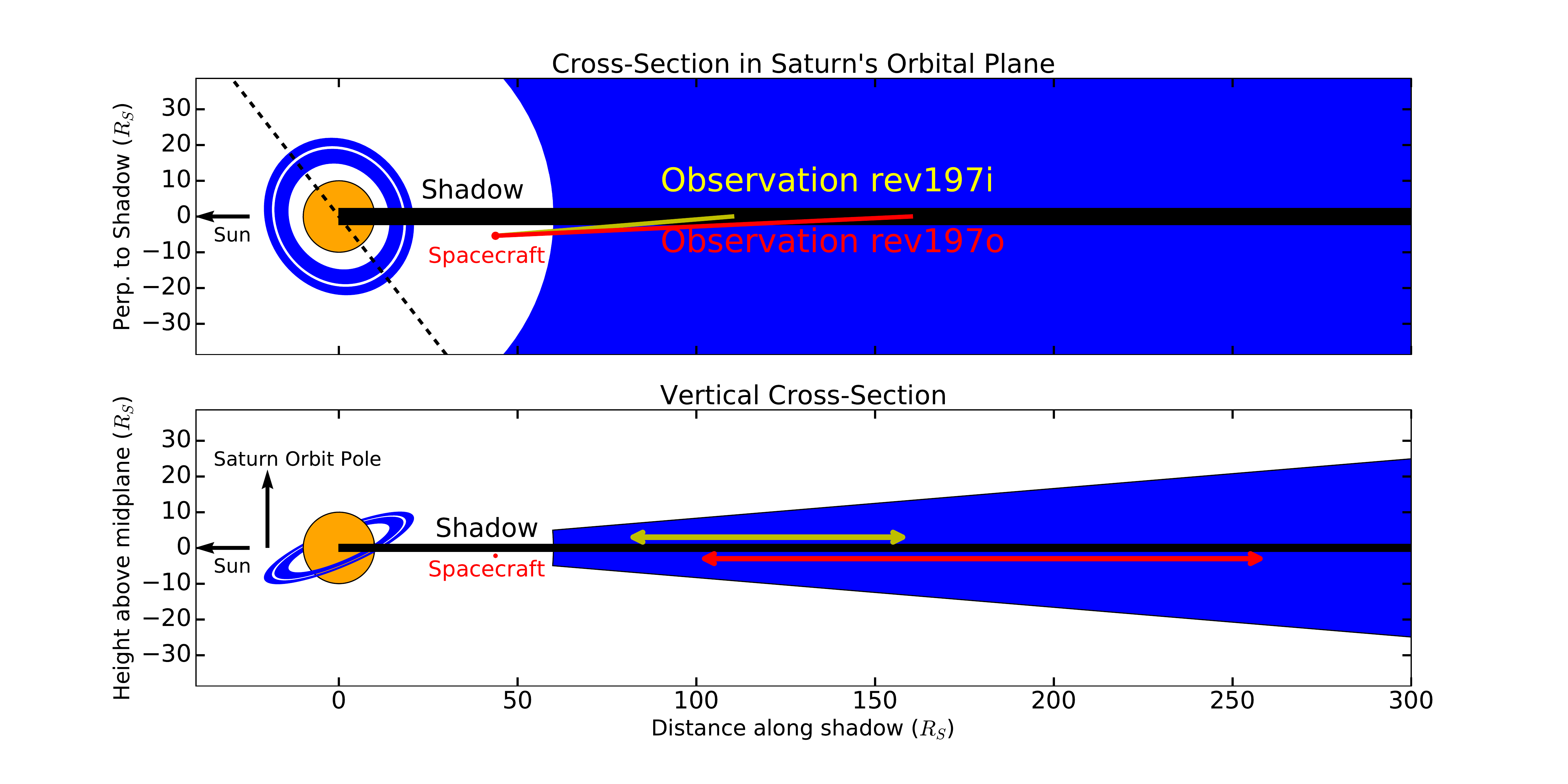}}
\caption{\label{geom} Sunlight enters from the left, and Saturn casts a shadow (black rectangle) extending to the right.
The spacecraft is plotted as a red circle, along with lines of sight to the center of the field of view for our inner (rev197i) and outer (rev197o) observations (described below in more detail).  
The top panel represents a cross-section along Saturn's orbital plane, through which Saturn's shadow passes, and which corresponds to the Phoebe ring's midplane.  
The bottom panel shows a vertical cross-section along the plane defined by the planet's shadow and its orbit pole, as well as color-coded double arrows denoting the radial extent spanned by each observation.
All distances are to scale, except for Saturn and its rings, which have been expanded by a factor of 10 to highlight their misalignment.
In both panels, the shadows show the actual size of Saturn and its rings.
The black dashed line in the top panel shows the intersection between Saturn's orbital and equatorial planes.
   }  
\end{figure*}

\begin{figure*}
\centering \resizebox{0.99\textwidth}{!}{\includegraphics{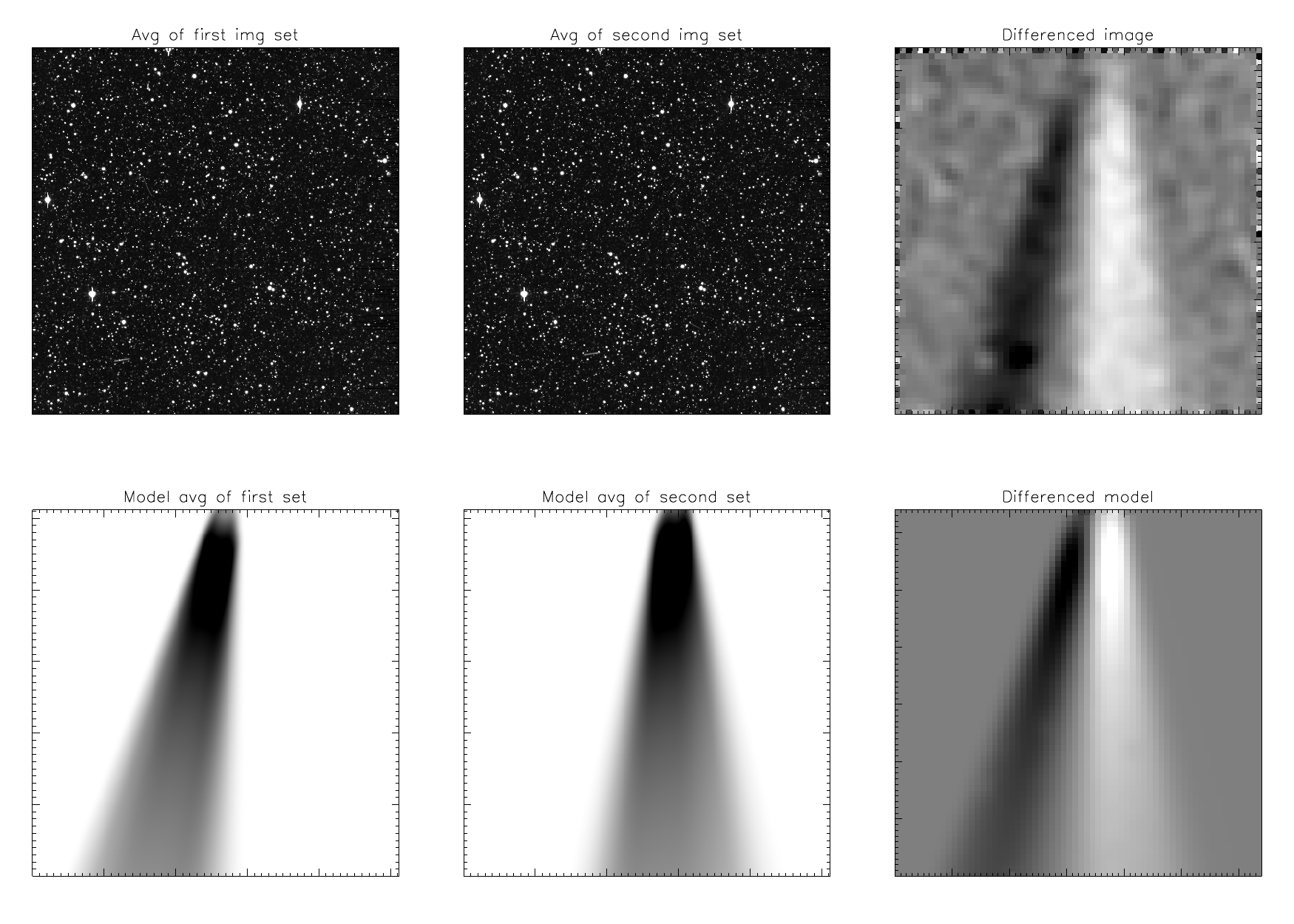}}
\caption{\label{4-sub}  Top left and top middle panels show averages of the first 25 and second 25 images in the rev197o pointing (with grayscale ranging from $I/F = [0,10^{-6}]$).
The corresponding panels below show the modeled dip in brightness along Saturn's shadow (range $= [-1.4\times 10^{-9}, 0]$, i.e., three orders of magnitude smaller than above).  
Subtracting the two average images (top right panel), attenuates the background signal while retaining the shadow signature (modeled in the bottom right panel).  To obtain the top right panel, we also filtered out noisy pixels, rebinned and smoothed the data.  The I/F range in the rightmost panels is $[-1.4\times 10^{-9}, 1.4\times 10^{-9}]$, and the bright and dark spots in the bottom left of the top right panel are the differenced signature of the irregular satellite Siarnaq, which happened to be in the field of view.  The field of view spans distances of $\sim 90 R_S$ (bottom of each image) - $300 R_S$ (top) from Saturn. Our models in the bottom three frames assume a uniform ring and are described in detail in Sec.\:\ref{dataconstant}.}  
\end{figure*}

The geometry is summarized in Fig.\:\ref{geom}.  
Over the span of the observation, the spacecraft (red point) does not move appreciably on the scale of the figure, but enough for the shadow to move across a large fraction of the camera's $3.5'\times3.5'$ field of view (see Fig.\:\ref{4-sub} and accompanying details below).  
The bottom panel additionally shows the radial ranges spanned by each observation (material beyond these limits contributed to fewer than $10\%$ of pixels in each pointing).
The shadow is wider in the top panel due to shadowing by the rings.
We also note that the depicted model for the Phoebe ring is simplified---it has been cut off at the orbital distance of Iapetus, which should intercept most of the material \citep{Tamayo11}, and it is drawn as symmetric about Saturn's orbital plane.
In reality, the Phoebe ring should begin warping toward Saturn's equatorial plane in the innermost regions of the disk (see Sec.\:\ref{dynamics}).

The corresponding observations for the outer section of the Phoebe ring (rev197o) are shown in Fig.\:\ref{4-sub}.

These observations (rev197o) comprise 50 220-second WAC exposures\footnote{Image names $W1758855456\_1 
 ,W1758855824\_1 
 ,W1758856192\_1 
 ,W1758856560\_1 
 ,W1758856928\_1 
 ,W1758857296\_1 \\
 ,W1758857664\_1 
 ,W1758858032\_1
 ,W1758858400\_1 
 ,W1758858768\_1 
 ,W1758859136\_1 
 ,W1758859504\_1 
 ,W1758859872\_1 \\
 ,W1758860240\_1 
 ,W1758860608\_1 
 ,W1758860976\_1 
 ,W1758861344\_1 
 ,W1758861712\_1 
 ,W1758862080\_1 
 ,W1758862448\_1 \\
 ,W1758862816\_1 
 ,W1758863184\_1 
 ,W1758863552\_1 
 ,W1758863920\_1 
 ,W1758864288\_1 
 ,W1758878396\_1 
 ,W1758878764\_1 \\
 ,W1758879132\_1 
 ,W1758879500\_1 
 ,W1758879868\_1 
 ,W1758880236\_1 
 ,W1758880604\_1 
 ,W1758880972\_1 
 ,W1758881340\_1 \\
 ,W1758881708\_1 
 ,W1758882076\_1 
 ,W1758882444\_1 
 ,W1758882812\_1 
 ,W1758883180\_1 
 ,W1758883548\_1 
 ,W1758883916\_1 \\
 ,W1758884284\_1 
 ,W1758884652\_1 
 ,W1758885020\_1 
 ,W1758885388\_1 
 ,W1758885756\_1 
 ,W1758886124\_1 
 ,W1758886492\_1 \\
 ,W1758886860\_1 
 ,W1758887228\_1.$} (using the clear filter CL1), centered on a point in the Phoebe ring 160 $R_S$ from the planet, at right ascension (RA) $= 223.7^{\circ}$, declination (Dec) = $-13.5^{\circ}$.  
In addition to the observations shown in Fig.\:\ref{4-sub} we obtained 47 exposures\footnote{Image names $W1758887712_1  
 ,W1758888080\_1  
 ,W1758888448\_1  
 ,W1758888816\_1  
 ,W1758889184\_1  
 ,W1758889552\_1  \\
 ,W1758889920\_1  
 ,W1758890288\_1  
 ,W1758890656\_1  
 ,W1758891024\_1  
 ,W1758891392\_1  
 ,W1758891760\_1  
 ,W1758892128\_1  \\
 ,W1758892496\_1  
 ,W1758892864\_1  
 ,W1758893232\_1  
 ,W1758893600\_1  
 ,W1758893968\_1  
 ,W1758894336\_1  
 ,W1758894704\_1  \\
 ,W1758895072\_1  
 ,W1758895440\_1  
 ,W1758895808\_1  
 ,W1758896176\_1  
 ,W1758896544\_1  
 ,W1758910652\_1  
 ,W1758911020\_1  \\
 ,W1758911388\_1  
 ,W1758911756\_1  
 ,W1758912124\_1  
 ,W1758912492\_1  
 ,W1758912860\_1  
 ,W1758913228\_1  
 ,W1758913596\_1  \\
 ,W1758913964\_1  
 ,W1758914332\_1  
 ,W1758914700\_1  
 ,W1758915068\_1  
 ,W1758915436\_1  
 ,W1758915804\_1  
 ,W1758916172\_1  \\
 ,W1758916540\_1  
 ,W1758916908\_1  
 ,W1758917276\_1  
 ,W1758917644\_1  
 ,W1758918012\_1  
 ,W1758918380\_1.$}, centered on a location 110 $R_S$ from Saturn, at RA = $225.0^{\circ}$, Dec = $-14.6^{\circ}$.  
We denote this data set further `inward' rev197i.  
The total observation window spanned 18 hours and 45 minutes, and time was evenly split between rev197o and rev197i.  
We collected all images in $2 \times 2$ summation mode due to data-volume constraints, and calibrated them with the standard Cassini ISS Calibration (CISSCAL) routines \citep{Porco04, West10} to apply flat-field corrections and convert the raw data to values of I/F, a standard measure of reflectance.
 
Following the techniques described in THB14, we first removed faulty pixels from the analysis, as well as ones with an I/F greater than a cutoff of $8\times10^{-8}$.  
Additionally, we found that removing particular images from the analysis improved our fits, due to offsets in the background levels between images at the level of our signal. 
To quantitively decide what images should be thrown out, we first calculated each image's mean brightness across the pixels that were not in shadow, as well as the standard error on each exposure's average I/F.  We then compared each image's mean value to the median across all exposures.  
If the deviation was greater than ten standard errors from the median, we discarded the image.

Finally, we employed the iterative procedure described in THB14 for removing cosmic rays and otherwise discrepant pixels, and applied a second iteration of image cuts as described above.  
In the end, our protocol retained $58\%$ of all pixels for analysis in both data sets (with 7 of the 50 images removed altogether in rev197o, and 5 out of 47 in rev197i).
This effectively removed the stars from the images and ensured a smooth background, allowing us to extract the faint Phoebe ring signal (this filtering process was used to obtain the right panels in Fig.\:\ref{4-sub}).

\subsection{Data Modeling} \label{dataconstant}

To quantitatively analyze the signal shown in Fig.\:\ref{4-sub}, we again adopted the procedure of THB14. 
This involved computing a simple shadow model (including a penumbra) for an oblate Saturn hosting completely opaque A and B rings.  
The shadow was correctly oriented and projected for the time of observation with the Navigation and Ancillary Information Facility (NAIF) SPICE toolkit \citep{Acton96} in each of the Cassini images.
We then calculated lines of sight for each pixel in the Cassini images through the shadow model, integrating each pixel's total pathlength through the shadow (see Fig. 1 in THB14).

To attenuate the constant background and extract the Phoebe ring signal, we generate a mean image of the $\sim 50$ exposures, and subtract this average from each of the images.
We then perform the same process on the set of modeled pathlength ``images."
Given their limited signal-to-noise-ratio data, THB14 assumed that there was a simple linear relationship between a pixel's I/F decrease and its corresponding line of sight's pathlength through the shadow.
This corresponds to a homogeneous Phoebe ring with constant dust-grain number density.

\begin{figure}[!ht]
\centering \resizebox{0.99\columnwidth}{!}{\includegraphics{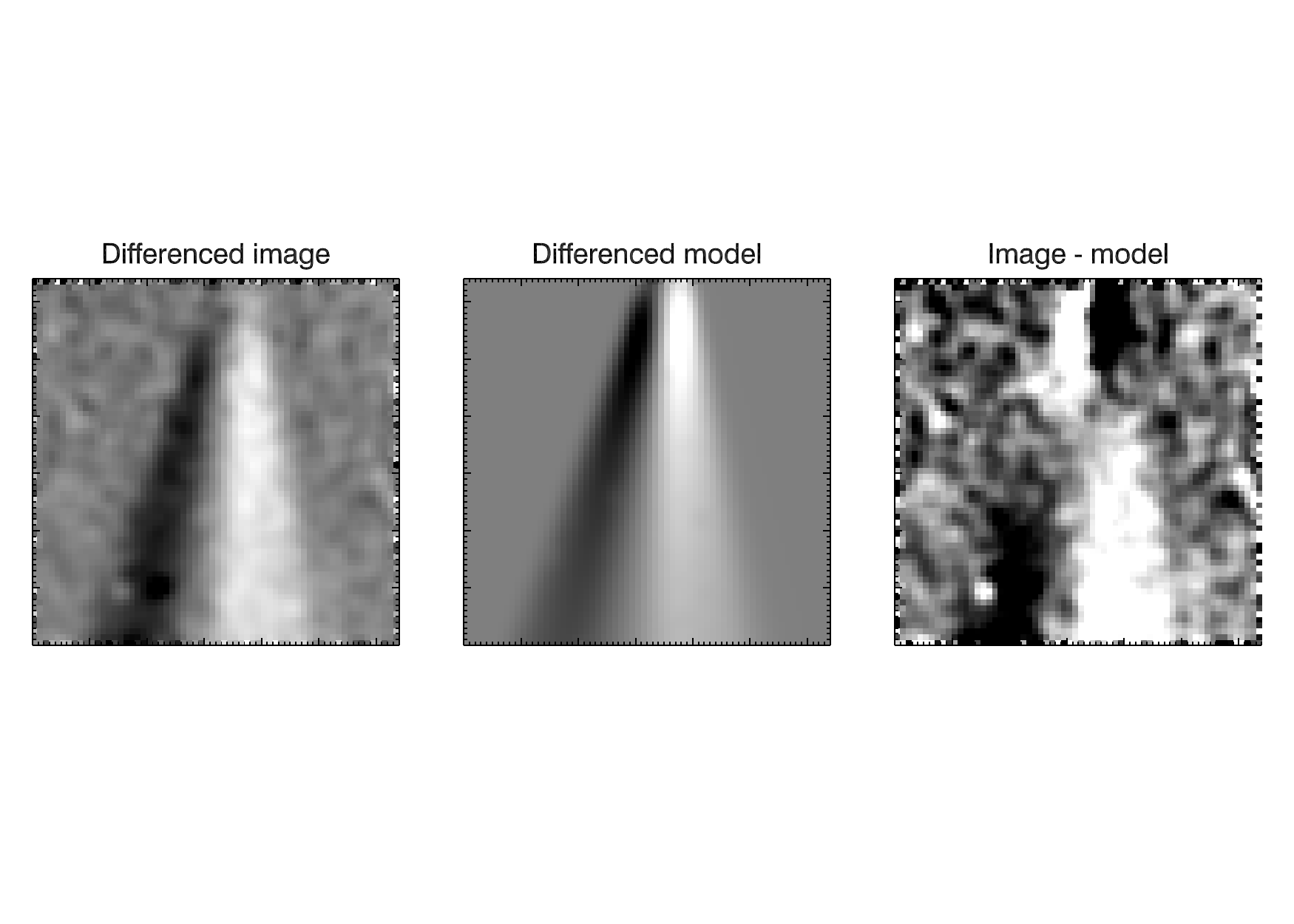}}
\caption{\label{4-constresid}  Left panel shows the real differenced data, corresponding to the top right panel of Fig.\:\ref{4-sub}.  
The middle panel shows the prediction assuming the best-fit homogeneous model for the Phoebe ring, and the right panel shows the result of subtracting the middle panel from the left one.  
The residuals suggest that there is substantial variation in the dust-grain number density as a function of distance from Saturn.
The color scale is the same as given in Fig.\:\ref{4-sub}.}  
\end{figure}

Figure \ref{4-constresid} shows our best-fit model when we similarly assume the Phoebe ring to be spatially homogeneous. 
While a constant-number-density Phoebe ring satisfactorily fit the noisier data of THB14, we see that our improved data deviate strongly from this model.
In addition, the pattern in the residuals in Fig.\:\ref{4-constresid} indicates that the Phoebe ring is fainter at increasing distance from the planet.
For this investigation, we therefore relax the homogeneity assumption and probe the Phoebe ring's radial structure (we assume there is no azimuthal variation across the shadow as the shadow's width represents less than 1\% of the ring's circumference).  
We note that one might expect such radial variation given the ring's expected radial extent $\sim 60-270 R_S$ \citep{Hamilton15}---if there were comparable amounts of material at different radii from Saturn, then the number density of particles would fall with distance as grains get spread over annuli of increasing volume.  

To model a radially varying Phoebe ring, we break it up into annuli that are each $10 R_S$ wide.  
We then assume that the I/F from the Phoebe ring in a given pixel is the sum of linear contributions proportional to the pathlengths through each of these annuli,
\begin{equation} \label{empirical}
I/F = \sum_i m_i p_i,
\end{equation}
where the sum runs over all the annuli, $p_i$ is the pathlength of the pixel's line of sight through the $i^{\text{th}}$ ring, and $m_i$ is the brightness per unit pathlength through the $i^{\text{th}}$ annulus.
We connect the $m_i$ to the physical size and spatial distributions of Phoebe ring grains in Sec.\:\ref{montecarlo}, but begin by obtaining empirical fits to the data using Eq.\:\ref{empirical}, assuming that the $m_i$ follow a power law with amplitude $A$ and power-law index $n$,
\begin{equation} \label{empiricalplaw}
m_i = A r_i^{n},
\end{equation}
where $r_i$ is the distance from Saturn to the middle of the $i^{\text{th}}$ annulus.

\begin{figure}[!ht]
\centering \resizebox{0.99\columnwidth}{!}{\includegraphics{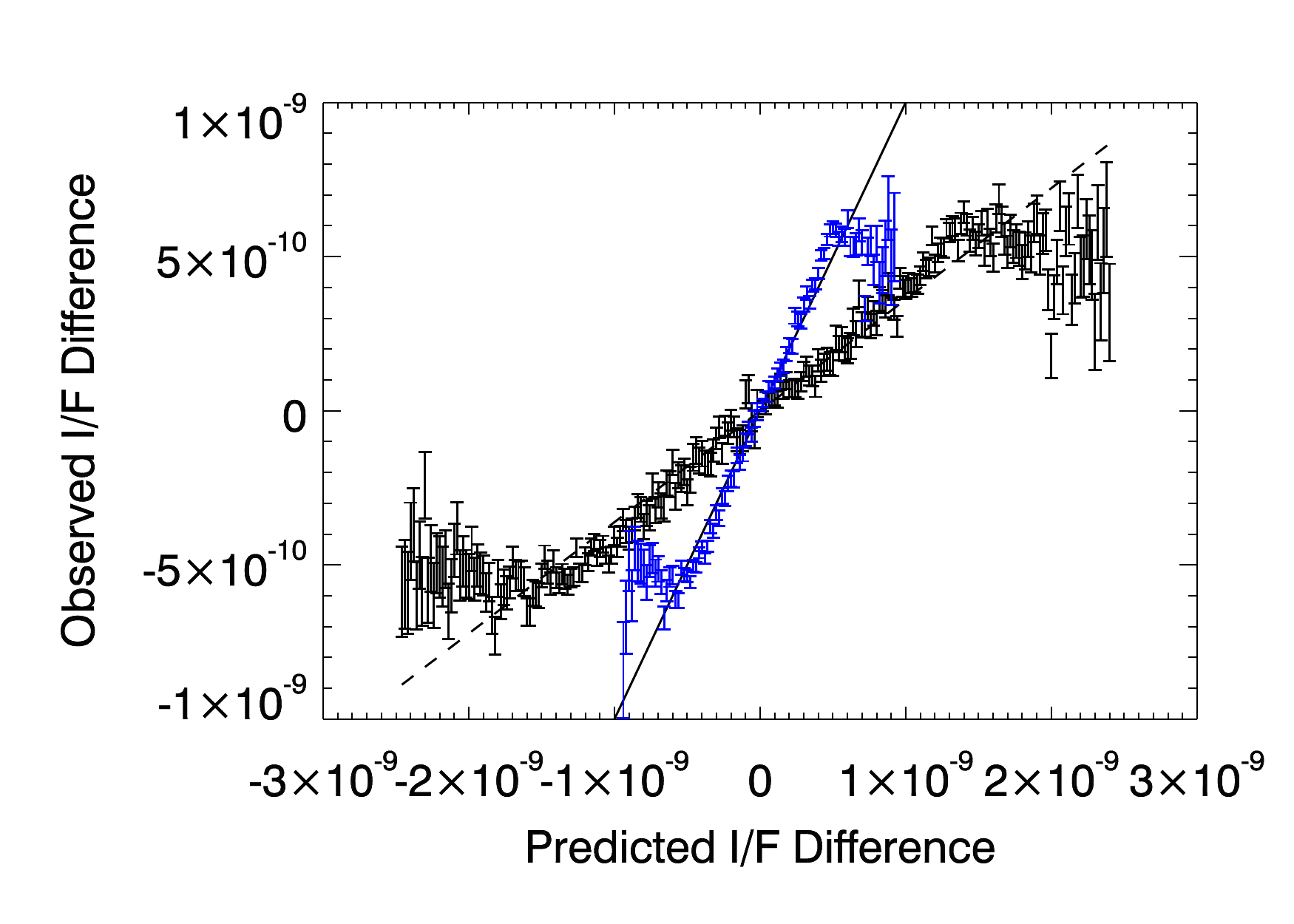}}
\caption{\label{fitline}  Observed vs. predicted I/F values (black).  
By fitting a line to these points (dashed), we can use the slope to correct $A$ (Eq.\:\ref{empiricalplaw}).  
This stretches/compresses the predicted I/F differences to match the expected slope of unity (blue points).  
However, while this model matches the overall slope, the fit deviates from a straight line at large predicted I/F differences due to a poor choice of the radial power-law index (Eq.\:\ref{empiricalplaw}).  
This model used $n = -0.5$.}  
\end{figure}

As above, we generate differenced observed and predicted images.
To quantitatively fit the data, we then bin all pixels by their predicted I/F values, calculate the mean predicted and observed I/F values in each bin, and estimate observed bin errors as the standard error $\sigma_i / N_i^{1/2}$, where $\sigma_i$ and $N_i$ are the standard deviation and number of pixels in bin $i$, respectively.
We then perform a least-squares fit to the line observed I/F = predicted I/F (solid black line in Fig.\:\ref{fitline}).

Because the model (Eq.\:\ref{empirical}) is linear in the amplitude $A$, we don't fit for it separately.
Instead, we guess an approximate amplitude for $A$ and first fit a straight line to the data (letting the slope vary, black line in Fig.\:\ref{fitline}).
We then divide our initial $A$ value by the best-fit slope, and recalculate predicted I/F values to obtain a line of unity slope (blue line, Fig.\:\ref{fitline}).
Of course, a wrong value of $n$ will still yield a bad fit (see the deviations at the ends of the lines), since the overall shape will deviate from a straight line even if the overall slope is approximately correct.  
This procedure to obtain the amplitude removes one of the fitted parameters, reducing the computational cost (for each model we predict and bin $\sim 10^7$ pixel values).
We tested that using this procedure and only fitting for $n$ consistently recovers the same models as when one fits for both parameters simultaneously.

\section{Results} \label{results}
\subsection{Single power-law}
We began by fitting models of the form Eq.\:\ref{empiricalplaw} for values of $n$ ranging from -5 to 3 with a step size of 0.125.
For the outer pointing (rev197o), we found a best-fit power-law index $n =-1.125$, which yielded a reduced $\chi^2$ of 1.18 with 111 degrees of freedom. 
For the inner pointing data (rev197i), we instead found a minimum $\chi^2$ at $n=-0.875$, corresponding to a reduced $\chi^2$ of 2.78 with 79 degrees of freedom\footnote{We find that the inner pointing consistently yields worse fits than the outer one. 
This may be due to complicated ring structure induced by the shifting equilibrium Laplace plane for small particles \citep{Tamayo13b, Rosengren14}, or by close encounters with Iapetus.
These processes are not captured by our simple empirical model, but we pursue them in Sec.\:\ref{dynamics}.}.  
Although this model substantially improved upon the assumption of a homogeneous Phoebe ring (reduced $\chi^2$ of 27.1 with 106 degrees of freedom), the lower slope and high reduced $\chi^2$ of the inner pointing indicate a single power law does not satisfactorily fit the data.
Upon further investigation, we found that the single power-law assumption tends to overpredict the I/F at small distances from Saturn ($\approx 100 R_S$). 
The compromise reached by the power-law fit thus tends to produce similar (though subdued) residuals to those for the homogeneous Phoebe ring model shown in Fig.\:\ref{4-constresid}.  

\subsection{Broken power-law} \label{brokenpowerlaw}
\begin{figure}[!ht]
 \centering \resizebox{0.99\columnwidth}{!}{\includegraphics{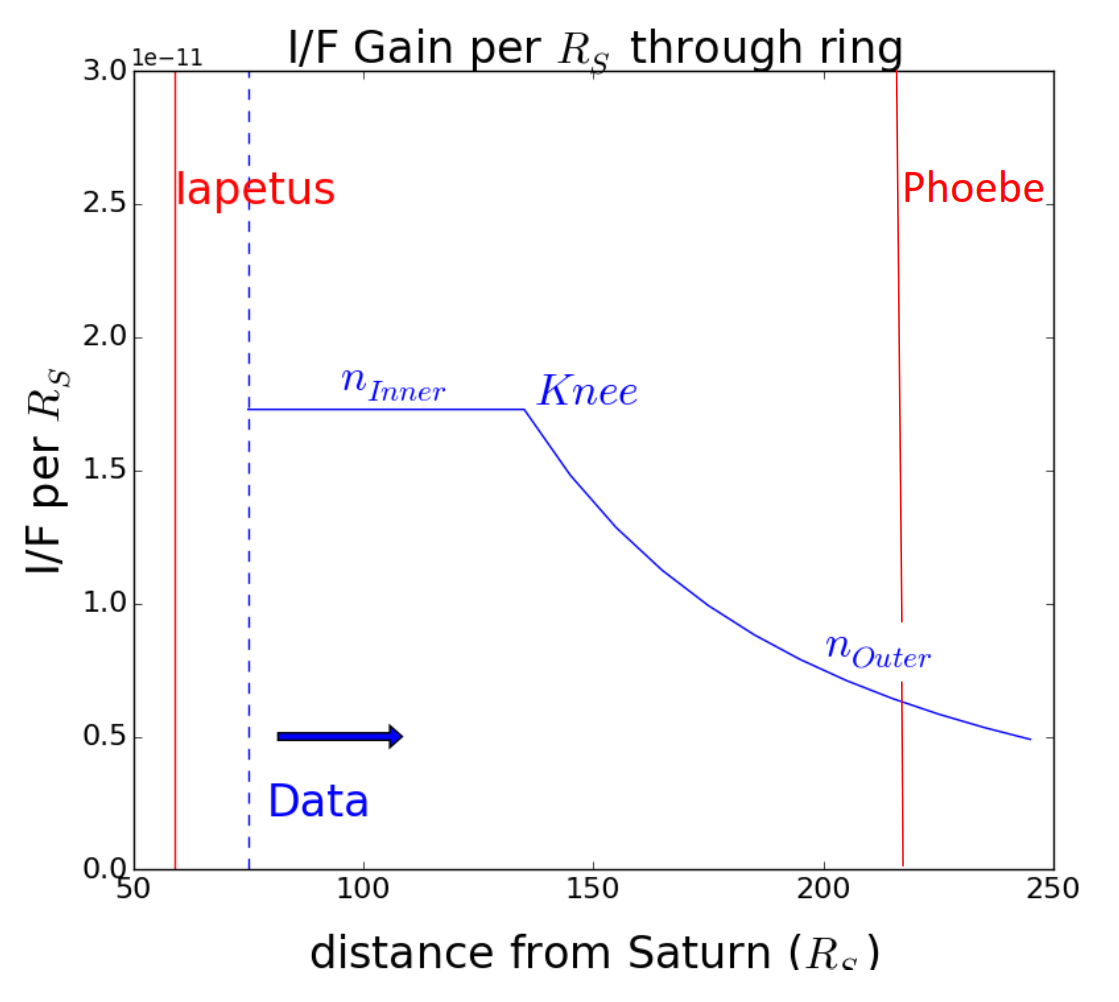}}
\caption{\label{knee-model}  Visualization of the three parameter broken power-law model, along with the semimajor axes of Iapetus and Phoebe, and the radial range captured in the data (right of dashed blue line).}
\end{figure}

In order to address our excess I/F prediction closer to Saturn, we then considered a broken power-law model,
\begin{eqnarray} \label{brokenplaweqs}
m_i(r_i < R_k) = A r_i^{n_{\text{Inner}}}, \\
m_i(r_i > R_k) = A r_i^{n_{\text{Outer}}},
\end{eqnarray}
where $R_k$ is the radial location of the `knee', where the power-law index shifts (see Fig.\:\ref{knee-model}).
Fitting the amplitude $A$ as discussed above, we now have three parameters, $n_{\text{Inner}}$, $n_{\text{Outer}}$ and $R_k$. 

We began by coarsely sampling a large section of parameter space.
Based on this initial investigation, we then settled on a finer grid of 2244 models, sampling $n_{\text{Inner}}$ from -3 to 5 in steps of 0.5, $n_{\text{Outer}}$ from -5 to 0 in steps of 0.25 and $R_k$ from 70$R_S$ to 180$R_S$ in steps of 10$R_S$. 
We found multiple local minima and dozens of models which offered compelling fits to the data (the global minimum reduced $\chi^2$ value was 0.993 and 1.665 for the outer and inner pointings, respectively). 

\begin{figure*}
\centering \resizebox{0.99\textwidth}{!}{\includegraphics{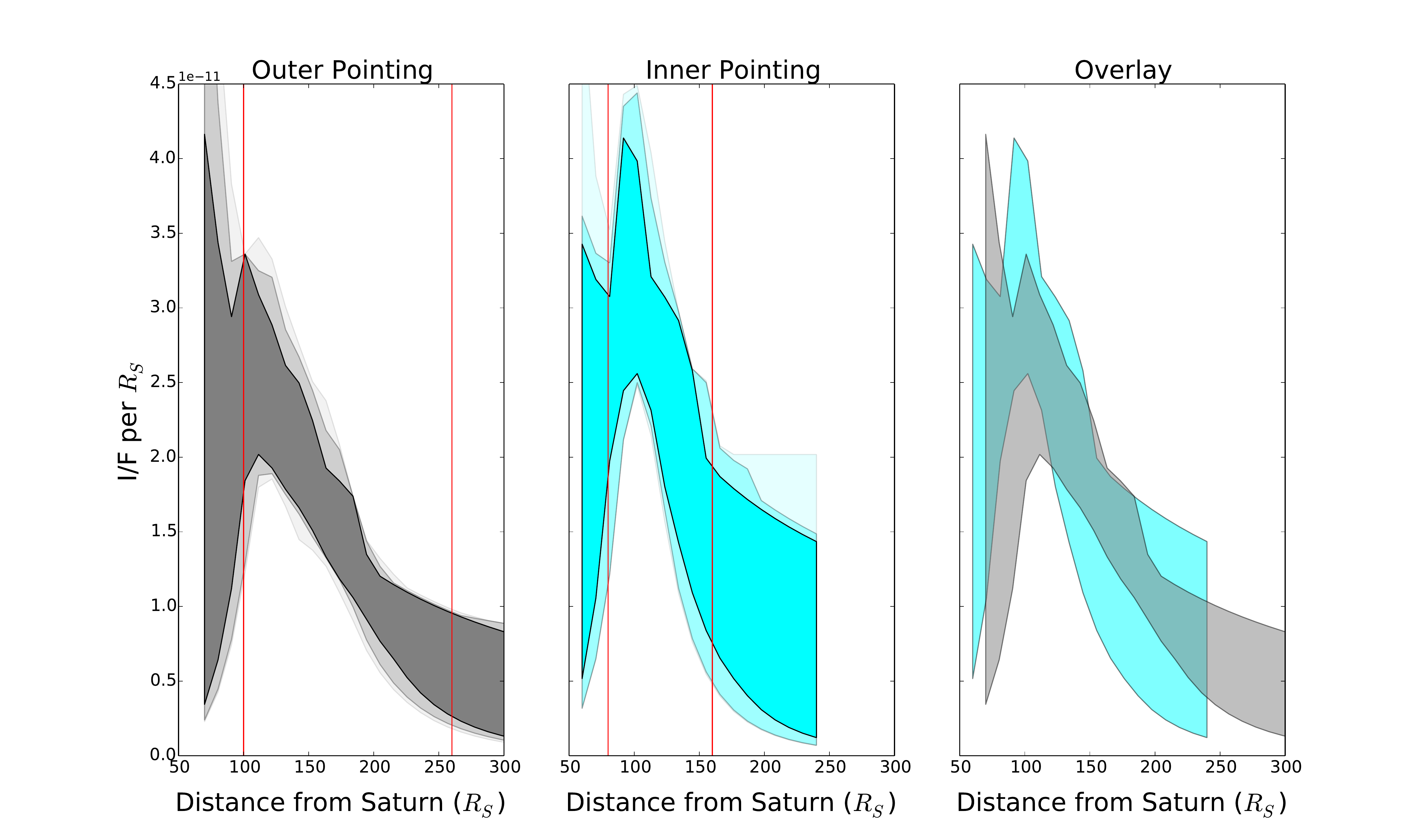}}
\caption{\label{best-vertline}  The best-fitting radial profiles for the Phoebe ring.
In each panel, the data mostly constrain the radial range between the two red lines (fewer than $10\%$ of pixels are influenced by Phoebe ring material lying outside this region).
The best-fit model's reduced $\chi^2$ was 0.99 and 1.67 for the outer and inner pointings, respectively.
Contours bound the models with reduced $\chi^2$ less than 1.25, 1.5 and 1.75 times the value for the best-fit model.
The right panel overlays the best contour for each of the two data sets.}
\end{figure*}

To check whether the best-fitting models indeed resembled each other, we graphed the various models and overplotted $\chi^2$ contours (Fig.\:\ref{best-vertline}).  All models with reduced $\chi^2$ within 25\% of the best-fit model ($\chi^2 = 0.99$ and 1.67 for the outer and inner pointing, respectively) lie within the darkest level surfaces, and additional contours are plotted at 1.5 and 1.75 times the minimum reduced $\chi^2$.  

The rapid increase in reduced $\chi^2$ across the contours of Fig.\:\ref{best-vertline} show that the data indeed constrains the Phoebe ring's radial profile.
As one would expect, the outer pointing (left panel) better constrains the radial profile at large distances, while the inner pointing (right panel) yields a narrower $\chi^2$ distribution close to Saturn.
Additionally, the darkest contours from the two panels in Fig.~\ref{best-vertline} show good agreement between the inner and outer pointing data.

The models (Fig.\:\ref{best-vertline}) suggest that the Phoebe ring's radial profile exhibits a steeper power-law decay at large radii that levels off closer in. 

\subsection{Alternate parametrizations of the data}
The range in models that fit the data well (Fig.\:\ref{best-vertline}) is a consequence of the experimental setup. 
In order to maximize the column of shadowed material along the line of sight, we performed the observations when Cassini was almost in Saturn's umbra (for these data, Cassini lay $\approx 5 R_S$ from the shadow axis, observing material $\sim 100 R_S$ away).  
Therefore, each pixel measures an integrated I/F deficit accumulated over a broad range of distances from Saturn.
This fundamentally limits the amount of radial information that can be extracted from the data. 

\begin{figure}[!ht]
 \centering \resizebox{0.99\columnwidth}{!}{\includegraphics{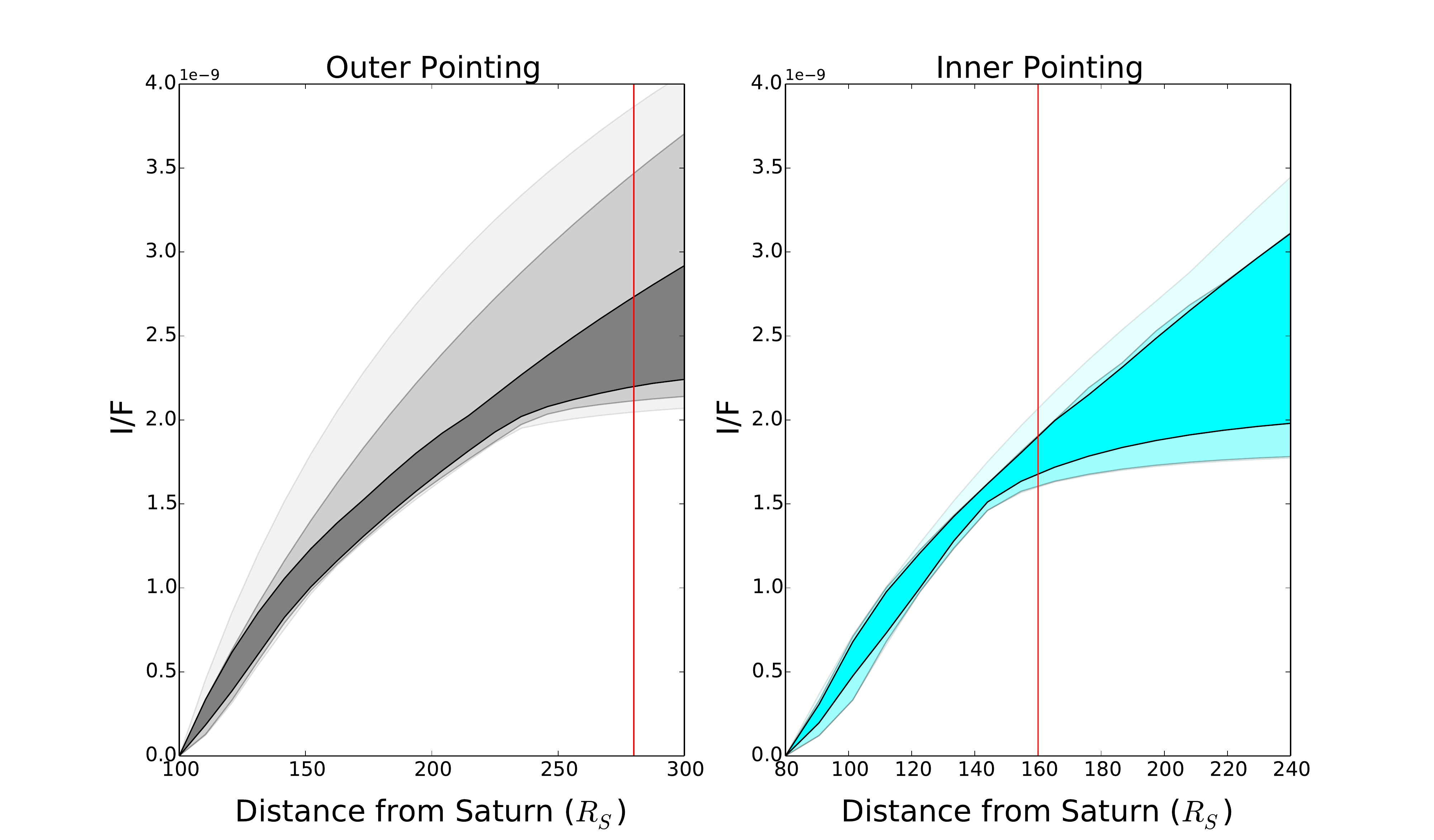}}
\caption{\label{integrals}  At each x-value, we integrate the models from Fig.\:\ref{best-vertline} from $80 R_S$(rev197i) or $100 R_S$(rev197o) to the radius in question.  
The contours show the spread in integrated I/F values for models that had reduced $\chi^2$ values in a given range (same contours as Fig.\:\ref{best-vertline}).  
The left gray panel corresponds to the outer pointing data, the right cyan panel corresponds to the inner pointing data. 
The inner endpoints to the integrations were chosen as the first radial slice that influenced at least $10\%$ of pixels in the dataset. 
The red line indicates the radius beyond which fewer than 10\% of pixels are affected by Phoebe ring material.}
\end{figure}

Alternatively, we can constrain the integrated I/F from the Phoebe ring looking (approximately radially) outward along the disk's midplane (in which the shadow lies).  
This can be approximated as the area under model curves plotted in Fig.\:\ref{best-vertline}.
As shown in the cumulative integrals of Fig.\:\ref{integrals}, we are better able to constrain this accumulated I/F than the individual contributions from separate radial slices. 
To show the range in models like in Fig.\:\ref{best-vertline}, we plot the same reduced $\chi^2$ contours.
In order to make these cumulative plots, we must choose an inner radius to begin the integral.
We chose the boundary radius where fewer than $10\%$ of pixels were influenced by Phoebe ring material lying inside this distance.
This corresponded to $80 R_S$ and $100 R_S$ in rev197i and rev197o, respectively (shown as the leftmost vertical red lines in the left and middle panels of Fig.\:\ref{best-vertline}).

We can obtain a simple estimate for the integrated I/F along the Phoebe ring's midplane by taking the mean integrated value across the models in the darkest contours of Fig.\:\ref{integrals}.
While it is difficult to calculate rigorous error bars for our measurements (see Appendix A), we choose to estimate the errors by taking the boundaries of the second contour shown in Fig.\:\ref{best-vertline}, which encompasses models that had a reduced $\chi^2$ less than 1.5 times the best-fit model's value.
This should be a conservative estimate given the large number of degrees of freedom (see Appendix A) and the fact that the reduced $\chi^2$ contours rise steeply beyond this contour (compare the second to third contours in Figs.\:\ref{best-vertline} and \ref{integrals}).  
We find\footnote{We took the values from the inner pointing data in the range $[80,100] R_S$, and those from the outer pointing over $[100,250] R_S$, adding the errors in quadrature.} that the Phoebe ring's integrated I/F from 80 to 250 $R_S$ along the disk's midplane is $2.7^{+0.9}_{-0.3} \times 10^{-9}$.

To connect these empirical fits to physical ring parameters, we now consider a dynamical model for Phoebe ring grains.

\section{A Dynamical Model for the Phoebe Ring} \label{dynamics}
The observational data indicate that the Phoebe ring's density does not decline with distance from the planet in a uniform way. 
Instead, something happens interior to 110 $R_S$ that causes its brightness profile to become significantly flatter.

One possibility is that this feature is due to Iapetus sweeping up material, as is theoretically expected \citep{Tosi10, Tamayo11}.
While the observed flattening of the Phoebe ring's radial profile occurs at roughly twice the orbital distance of Iapetus, the moon may be intercepting grains on orbits that are rendered highly eccentric by radiation pressure.
An alternate possibility are instabilities in the equilibrium Laplace surface, which governs Phoebe ring grains' vertical orbital evolution.
\cite{Rosengren14} found that for certain grain sizes, local Laplace equilibria can become unstable as grains evolve inward, forcing them to suddenly oscillate around a newfound, distant equilibrium.  
A ring composed of single-sized dust grains in this range would therefore puff up at a characteristic distance from Saturn, while a distribution of particle sizes in this range would grow vertically more gradually.
This effect would thus also tend to decrease the density of dust particles in Saturn's shadow, and thus the observed brightnesses.

In order to assess the above possibilities, and interpret our observational results, one must construct a dynamical model for Phoebe ring grains.
We begin by assessing the 3-dimensional geometry involved.

\begin{figure}[!ht]
 \centering \resizebox{0.99\columnwidth}{!}{\includegraphics{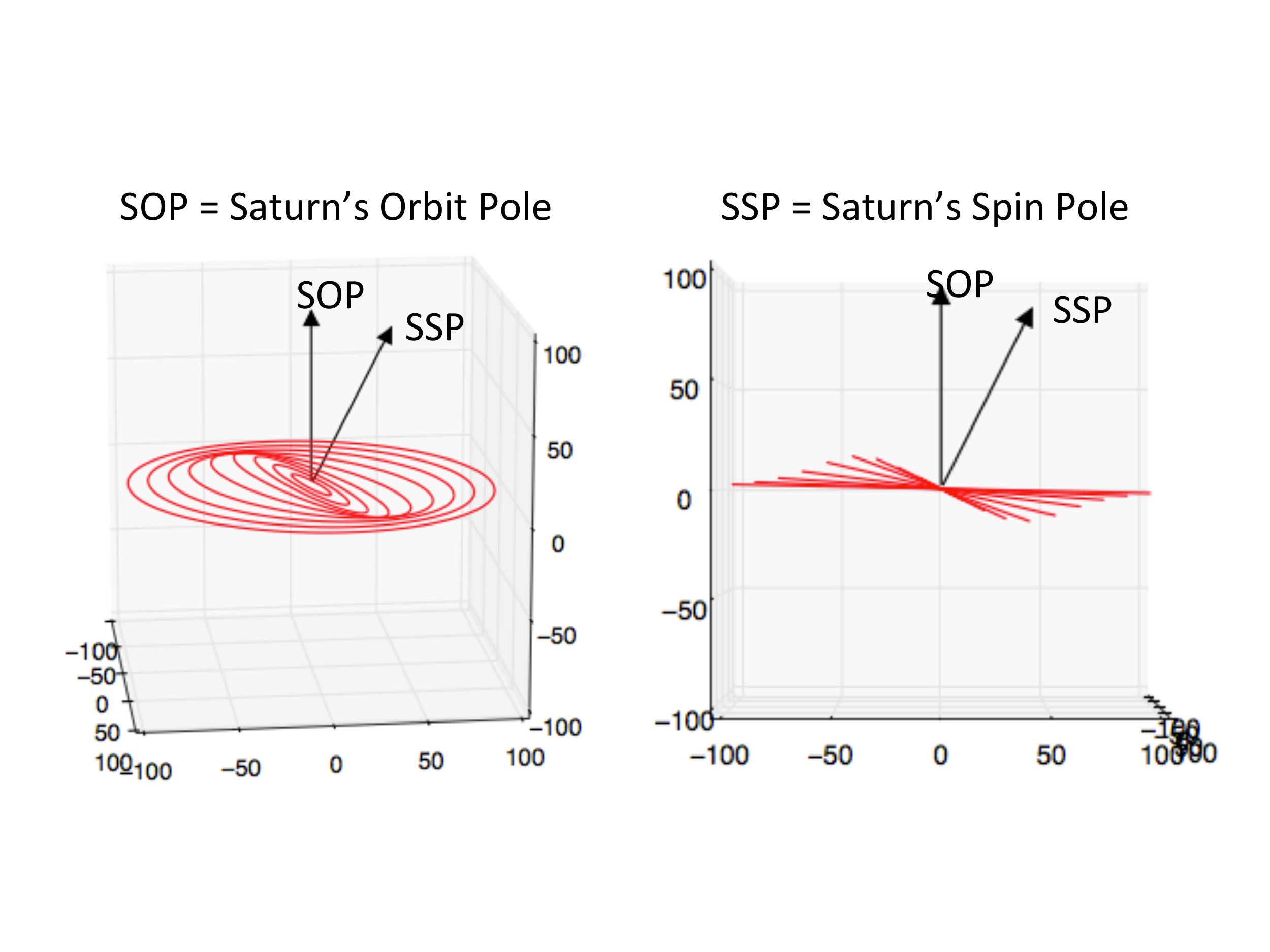}}
\caption{\label{4-warped}  On the left is an oblique view from above of the warped equilibrium surface that an infinitely thin Phoebe ring would follow.  
On the right is an edge-on view.  
The horizontal plane in the edge-on view corresponds to Saturn's orbital plane, which is the plane in which Saturn casts its shadow.  
If Phoebe-ring material were spread evenly from Phoebe to the central planet, our method would observe an inner edge to the ring due to material tilting off the plane that is probed by Saturn's shadow.  
The only exception would be if the Sun happened to be aligned with the line along which Saturn's orbit plane intersects its equatorial plane (vantage point shown in right panel)---along this line, material would extend inward to the planet; however, in our data, the Sun lies at about $49^{\circ}$ from this line of nodes.  
All distances are in Saturn radii.}
\end{figure}

The shadow cast by Saturn lies in the planet's orbital plane.  
This is the equilibrium orbital plane for particles far from Saturn, and thus the symmetry plane for the Phoebe ring at large distances, making our observations possible.
But as dust grain orbits decay inward, they will follow the local equilibrium plane, which gradually shifts toward Saturn's equatorial plane (see Sec.\:\ref{4-inc}).  
An infinitely thin Phoebe ring would therefore follow a warped surface like that shown in Fig.\:\ref{4-warped}.  
The real Phoebe ring has a thickness about the equilibrium surface that is set by the orbital inclination that particles inherit from Phoebe (the thickness increases linearly with distance from Saturn, to a value of $\approx 40 R_S$ at Phoebe's distance of $215 R_S$, \citealt{Verbiscer09}).
Because closer to the planet the ring lifts out of the plane probed by Saturn's shadow (the planet's orbital plane), observing an inner edge to the Phoebe ring using our technique does not necessarily point to Iapetus sweeping up material.  
Of course, in this case Iapetus might (and should) nevertheless be carving out an inner edge to the Phoebe ring; this would just occur in a region inaccessible to our observation technique.\footnote{Iapetus orbits in a plane intermediate between Saturn's equatorial and orbital planes, but would nevertheless sweep up the material as the dust grains' and moons mutually precess into configurations where collisions are possible \citep[see][]{Tamayo11}.}

In order to address this issue, we now consider the dynamics of particle orbits as they decay inwards toward Saturn.  
We will use these results below to generate a Monte Carlo simulation of the Phoebe ring's 3-D structure.  
In particular, we study the evolution of dust-grain orbits under the simultaneous influence of radiation pressure, tidal solar gravity, and Saturn's oblateness.  
Note that we are ignoring the gravity of Iapetus, which could alter the dynamics at semimajor axes where the orbital periods of the particle and Iapetus form a near-integer ratio.  
In addition, we approximate Saturn's orbit as circular.

There are many possible trajectories depending on grains' initial conditions, their physical radius, and the position of the Sun at the time of their launch.
To try and circumvent the computational cost associated with this large phase space, we develop an analytical model for the dynamics, which we then sample from at random times to build up the 3-D structure of the Phoebe ring.
To make analytic progress, we assume that the evolutions of the eccentricity and the inclination are decoupled.  
In particular, we calculate the inclination evolution assuming a circular orbit, and evaluate the eccentricity evolution assuming a planar orbit.  
This amounts to ignoring second-order eccentricity terms in the equations of motion for the inclination evolution and vice-versa.  
These assumptions are only rigorously correct for circular orbits around a planet with zero obliquity (so that all perturbations act in the same plane); however, they are reasonable approximations as long as the planet's obliquity is not too large (Saturn's obliquity $\approx 26.7^{\circ}$) and the orbital eccentricities are moderate.  
We will compare our analytic results to direct integrations below.

\subsection{Inclination Evolution} \label{4-inc}
Because, under the perturbations stated above, the inclination evolution is slow compared to both the particle's orbital timescale around Saturn and Saturn's orbital period about the Sun, one can profitably average over these fast oscillations.  
When considering only the effects of the quadrupole potentials from the planet's oblateness and the Sun's gravity, one then obtains the classical result that, for a given circumplanetary orbit's semimajor axis, an equilibrium plane exists between the planet's equatorial and orbital planes.  
Particle orbits in this so-called {\it Laplace plane} remain in the plane (in this sense it is an equilibrium plane), whereas inclined orbits will precess around the Laplace plane normal at approximately constant inclination.  
The local Laplace plane represents a compromise between the oblateness perturbations that dominate close to the host planet and are symmetric about its equatorial plane, and the solar perturbations that dominate far out and are symmetric about the planet's orbital plane.  
Thus, the local Laplace plane nearly coincides with the planet's orbital plane for distant particle orbits (e.g., those of the irregular satellites), while progressively smaller orbits have their respective Laplace planes transition toward the planet's equatorial plane (see Fig.\:\ref{4-warped}).  
The shift between these configurations occurs at approximately the Laplace radius $r_L$, where the torques from the two perturbations roughly balance \citep{Goldreich66}, or

\begin{equation} \label{rl}
{r_L}^5 \approx 2 J_2 {R_p}^2 {a_p}^3 (1 - e_p^2)^{3/2} \frac{M_p}{M_{\odot}},
\end{equation}
where $J_2$ is the quadrupole coefficient from an axisymmetric expansion of the planet's gravitational potential, $R_p, M_p, a_p$ and $e_p$ are the planet's radius, mass, orbital semimajor axis and eccentricity, respectively, and $M_\odot$ is the Sun's mass.
Considering the contribution of the inner satellites to Saturn's effective $J_2$, $r_L$ at Saturn is $\approx 55 R_S$ \citep{Tremaine09}; Iapetus, at $59 R_S$, has a mean orbital plane set by the local Laplace plane's inclination to Saturn's orbital plane of $\approx 11.5^{\circ}$.

The inclusion of radiation pressure (which is symmetric about the planet's orbital plane, like solar tides) shifts the balance between the planet's oblateness and solar gravity.  
Because radiation-pressure-induced precession of retrograde orbits opposes solar gravity precession, this is equivalent to a weakened effective solar gravity.  
Thus, for retrograde orbits, the transition of the equilibrium plane from the planet's orbital to equatorial planes occurs outside the classical Laplace radius given by Eq.\:\ref{rl}.  
Conversely, radiation pressure enhances the solar-gravity-induced precession of prograde orbits, so the transition radius moves inward \citep{Tamayo13b}.  
Additionally, because radiation pressure is particle-size dependent, the local Laplace planes for grains of different sizes will vary.

Recently, \cite{Rosengren14} performed a rigorous analysis of Laplace plane equilibria modified by radiation pressure.  
In Sec.\:\ref{montecarlo} we use their Eq.\:32 to calculate the equilibrium Laplace plane orientation for a given particle orbit's semimajor axis and particle radius.
For initial orbital orientations outside the corresponding equilibrium plane, we assume uniform precession about the Laplace plane at constant inclination.

\subsection{Eccentricity Evolution} \label{4-ecc}
We now consider the evolution of the orbital eccentricity, assuming a planar orbit around a planet with zero obliquity (we will compare our results to direct integrations with a tilted Saturn below).  
As opposed to the orbital inclination, the orbital eccentricity of small grains will undergo large-amplitude oscillations over a single Saturn year \citep{Burns79}.  
Moreover, such retrograde particle orbits that begin on near-circular orbits will reach their maximum eccentricities when their pericenter is aligned with Saturn's shadow (where we make our observations).  
It is therefore important not to average over Saturn's orbit about the Sun (period $\approx 30$ yrs) in this application (but we still average over the much faster particle orbit around Saturn, which has a period of $\sim 1$ year).
\cite{Hamilton96} have studied this problem for prograde orbits.  We now take their prograde solutions and apply symmetry arguments to derive the equations of motion for retrograde orbits.  

In a frame centered on Saturn, the equations of motion for a prograde orbit can be written as 
\begin{eqnarray} \label{4-eom}
\frac{1}{n_\odot}\frac{d\varpi}{dt} &=& A \sqrt{1-e^2}[1 + 5 \cos{2(\varpi - \lambda_\odot)}] + C \frac{\sqrt{1-e^2}}{e} \cos{(\varpi - \lambda_\odot)} + \frac{W}{(\sqrt{1-e^2})^2}, \nonumber \\ 
\frac{1}{n_\odot}\frac{de}{dt} &=& 5Ae\sqrt{1-e^2} \sin{2(\varpi - \lambda_\odot)} + C\sqrt{1-e^2} \sin{(\varpi - \lambda_\odot)},
\end{eqnarray}
where $e$ is the particle's orbital eccentricity, $\varpi$ is the longitude of the grain orbit's pericenter, $\lambda_\odot$ is the longitude of the Sun as it ``orbits" around Saturn in the saturnocentric frame, $n_\odot$ is the Sun's angular rate, and $A$, $C$ and $W$ are dimensionless constants capturing the strength of the sun's tidal gravity, radiation pressure, and the planet's oblateness, respectively:
\begin{equation}
A \equiv \frac{3}{4}\frac{n_\odot}{n}, \;\;\;\;\; C \equiv \frac{3}{2} \frac{n}{n_\odot}\sigma, \;\;\;\;\;\; W \equiv \frac{3}{2} J_2 \Bigg(\frac{R_p}{a}\Bigg)^2 \frac{n}{n_\odot},
\end{equation}
where $n$ is the particle's mean motion, $a$ is the particle orbit's semimajor axis, and $\sigma$ is the ratio of the radiation pressure force to the gravitational force of the planet on the body at a distance $a$ \citep[see Eq. 3 of][]{Hamilton96}.

A retrograde orbit is prograde in a frame where time runs backward, so we can immediately write down the equations of motion from Eq.\:\ref{4-eom} in this flipped frame.  
To be explicit, we can write $dt$ as $dt^-$ to emphasize that it represents time in the flipped frame, and we can then obtain the equations of motion for a retrograde orbit in a frame where time runs forward by re-expressing the equations of motion in terms of the original variable $t$ (through the simple relation $dt^-$ = $-dt$).  
Note that in applications with non-zero inclination one must be careful to also write $i^-$, $\varpi^-$, etc., when applying the prograde equations of motion and then re-express these in terms of $i$, $\varpi$, etc.  
This is because when flipping the time, $i \rightarrow 180-i$, the ascending node changes by $180^{\circ}$, and $\varpi = \Omega + \omega \rightarrow \Omega - \omega$.  

Additionally, one might be tempted to write $\lambda_\odot = n_\odot t$ in Eq.\:\ref{4-eom}, and have it flip sign upon these transformations; however, in the flipped frame the Sun moves backwards at a rate $- n_\odot t$, so one would write the equations in the flipped frame with terms involving, $\varpi + n_\odot t^-$, which would revert to $\varpi - n_\odot t$ when re-expressed in the original frame.  
Physically, the relevant terms in the differential equations only depend on the instantaneous position of the Sun, $\lambda_\odot$, not the direction in which it is moving, which is why we chose to express the right-hand sides of Eq.\:\ref{4-eom} in terms of $\lambda_\odot$.  

The above steps yield retrograde equations of motion with the signs on the right-hand sides of Eq.\:\ref{4-eom} negated.  
Following \cite{Hamilton96}, we now move to a frame where the $x$ axis rotates with the Sun at a rate $n_\odot t$ so that the potential is stationary, as this will yield a conserved quantity.  
Note that this would not be strictly true for a planet on an eccentric orbit (as the Sun would no longer ``move" at a constant rate), or if the obliquity were nonzero (as the oblateness potential would become time-dependent).  
Denoting the longitude of pericenter relative to the Sun's position $\phi_\odot = \varpi - \lambda_\odot$, and plugging in for $d\varpi/dt$ from Eq.\:\ref{4-eom}, we have the equations of motion for a retrograde orbit,
\begin{eqnarray} \label{4-eomrot}
\frac{1}{n_\odot}\frac{d\phi_\odot}{dt} &=& \frac{1}{n_\odot}\frac{d\varpi}{dt} - 1 \nonumber \\
&=& -A \sqrt{1-e^2}[1 + 5 \cos{2(\phi_\odot)}] - \nonumber \\
&&C \frac{\sqrt{1-e^2}}{e} \cos{(\phi_\odot)} - \frac{W}{(\sqrt{1-e^2})^2} - 1, \nonumber \\ 
\frac{1}{n_\odot}\frac{de}{dt} &=& -5Ae\sqrt{1-e^2} \sin{2(\phi_\odot)} -C\sqrt{1-e^2} \sin{(\phi_\odot)}. \nonumber \\
\end{eqnarray}

Following \cite{Hamilton96}, we can write these equations of motion using 
\begin{equation}
\frac{1}{n_\odot}\frac{de}{dt} = - \frac{\sqrt{1-e^2}}{e} \frac{\partial{\mathcal{H}}}{\partial{\phi_\odot}},\;\;\;\;\;\;\frac{1}{n_\odot}\frac{d\phi_\odot}{dt} =  \frac{\sqrt{1-e^2}}{e} \frac{\partial{\mathcal{H}}}{\partial{e}}
\end{equation}
and a conserved ``Hamiltonian"\footnote{$\mathcal{H}$ is not strictly a Hamiltonian since the equations of motion are not canonical.}
\begin{equation} \label{4-H}
\mathcal{H} = \sqrt{1-e^2} - \frac{1}{2} A e^2 [1 + 5\cos(2\phi_\odot)] - Ce\cos\phi_\odot - \frac{W}{3(1-e^2)^{3/2}};
\end{equation}
cf. Eq. 9 in \cite{Hamilton96}.  Trajectories in this one degree-of-freedom problem thus move on level curves of constant $\mathcal{H}$.

\begin{figure}[!ht]
 \centering \resizebox{0.99\columnwidth}{!}{\includegraphics{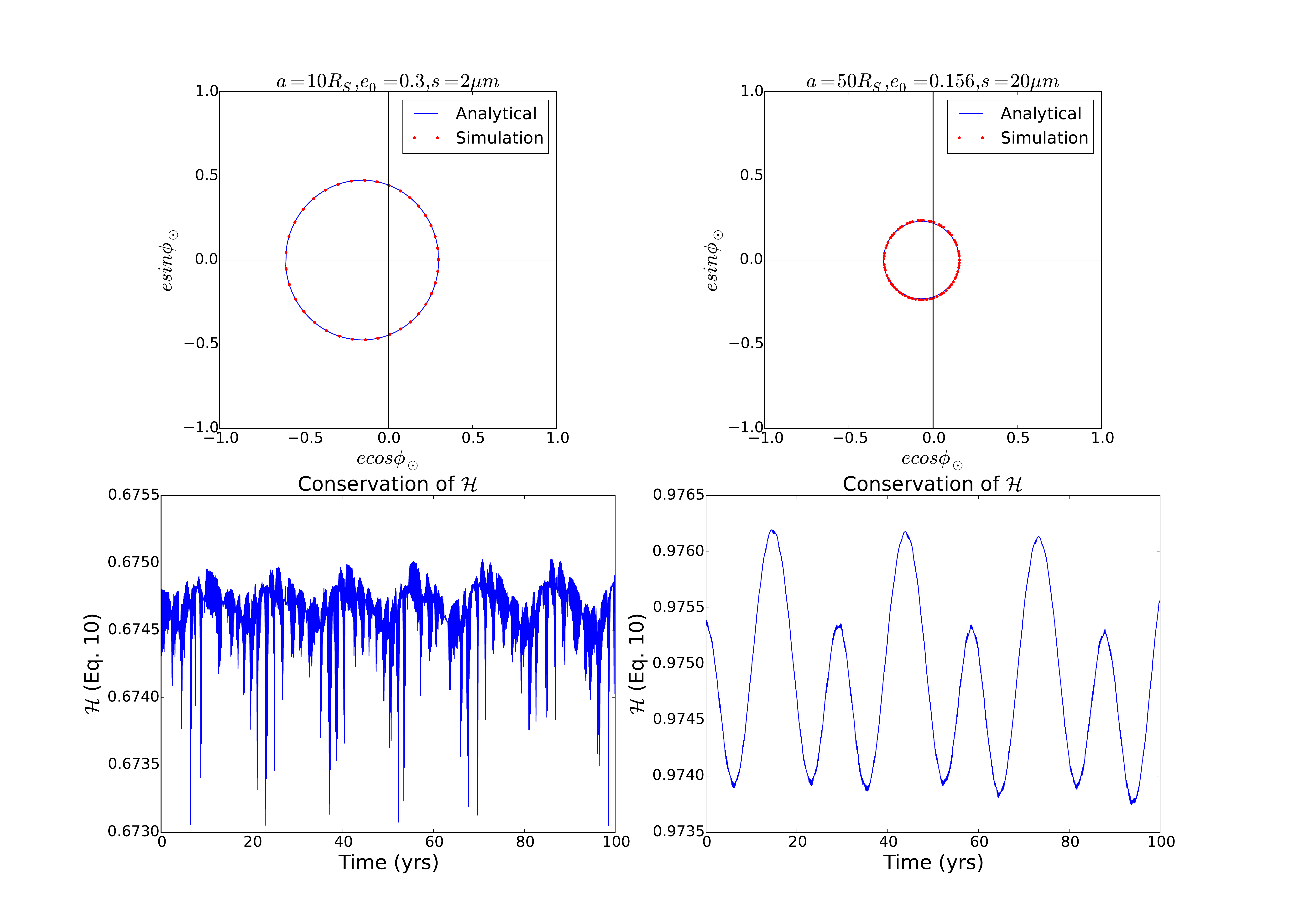}}
\caption{\label{4-intcomp} Top panels compare our analytical level curves (blue, calculated with Eq.\:\ref{4-H}, and thus assuming a zero-obliquity planet in a circular orbit) to direct integrations (red) with Saturn's present eccentricity and obliquity.  
See text for the parameters of the two integrations.
The bottom two panels show that the analytical $\mathcal{H}$ (Eq.\:\ref{4-H}) is conserved to within $1\%$ in the numerical simulations.}
\end{figure}

Figure \ref{4-intcomp} compares these results to two direct integrations that include Saturn's obliquity and its orbital eccentricity.  
The numerical integrations were performed with the well-established dust integrator \citep{Hamilton93, Hamilton08, Jontof12, Jontof12b}.
In both test cases, the particles were launched such that their orbits' pericenters coincided with the direction toward the Sun at $t=0$, and both simulations were run for 100 years (i.e., more than three Saturn orbits, and $\sim 7000$ particle orbits (left panels) and $\sim 400$ particle orbits (right panels). 
The top panels show the evolution of $e$ and $\phi_\odot$ in polar plots, where the radial distance gives the eccentricity, and the angle from the positive $x$ axis gives $\phi_\odot$.  
The top left panel shows a $2\mu$m grain on a retrograde orbit with $a=10R_S$ and initial eccentricity 0.3, in Saturn's equatorial plane (which is effectively coincident with the local Laplace plane at this semimajor axis).  
Despite the large eccentricities (reaching values greater than 0.5), the agreement is excellent.  
The bottom left panel shows the corresponding evolution of the analytical $\mathcal{H}$ (Eq.\:\ref{4-H}) in the numerical simulations, which we verify is conserved to well within $1\%$.  

The right panels are for a $20 \mu$m particle in an orbit with $a=50 R_S$, initial orbital eccentricity $0.156$ (the value for Phoebe), and initial orbital inclination to Saturn's orbital plane of $165^{\circ}$, roughly half way between the orbital and equatorial planes.  
These values were chosen as representative of the grains we wish to simulate, at points where our neglect of the coupling between eccentricity and inclination could be problematic;  Phoebe's eccentricity (which grains that are launched at slow speeds should inherit) is currently 0.156, and grains smaller than $\sim 3 \mu$m would be quickly eliminated by radiation pressure upon being liberated from Phoebe \citep{Verbiscer09, Tamayo11}.  
Again, the agreement is excellent, so we conclude that our analytic model should provide valuable insight into the orbits of most grains in the Phoebe ring size distribution.  
One should keep in mind, however, that the orbits of the smallest particles that survive immediate elimination by radiation pressure may exhibit important deviations from our results, particularly near and beyond the Laplace plane transition, where Saturn's oblateness no longer dominates.

\subsection{Monte Carlo Simulations} \label{montecarlo}
With an approximate analytical model in hand, we can efficiently generate a Monte Carlo simulation of the Phoebe ring where we randomly sample particle positions in their orbital evolution, and see how many particles lie in Saturn's shadow at various radial distances from the planet.  
But to compare these simulations with our observations, we must first connect our model to the photometry.

For low-optical-depth clouds like the Phoebe ring, the I/F scattered by ring particles is related to the line-of-sight optical depth $\tau$, the phase function $P(\alpha)$, where $\alpha$ is the phase angle, and the single-scattering albedo $\varpi_0$ (at 0.635 $\mu$m, where our observing band is centered, \citealt{Porco04}) through \citep{Burns01}
\begin{equation}
\frac{I}{F} = \frac{1}{4} \tau \varpi_0 P(\alpha).
\end{equation}
Writing $d\tau = n \sigma dl$, where $n$ is the number density of particles, $\sigma$ is their geometrical cross-section, and $dl$ is a differential length element along the line of sight, we define
\begin{equation} \label{IF}
m(r) \equiv \frac{d(I/F)}{dl} = \frac{n(r) \sigma \varpi_0 P(\alpha)}{4},
\end{equation}
The local quantity $m(r)$ thus quantifies how much $I/F$ is gained per differential pathlength through the Phoebe ring, and is the function we wish to extract from the observations.  
Since we only make measurements along Saturn's shadow, which subtends a small azimuthal angle, we take $m$ to only be a function of the distance from the planet, $r$.  
For a given model of $m(r)$, one obtains the expected change in $I/F$ in one of our pixels by integrating $m(r)$ along the path through the shadow.  

Since not only the cross-section, but also the number density (through the orbital dynamics discussed above) will be particle-size dependent, we generalize Eq.\:\ref{IF} to consider a range of particle sizes, obtaining the differential contribution to $m(r)$ from grains with radii between $s$ and $s + ds$,
\begin{equation}\label{dm}
dm(r,s) = \frac{\pi s^2 \varpi_0 P(\alpha)n(r,s)ds}{4},
\end{equation}
where $n(r,s)$, the differential number density for particles between size $s$ and $s + ds$ lying between $r$ and $r + dr$ from Saturn.  
Because the observation's wavelength (0.635 $\mu$m) is much shorter than even the smallest long-lived dust grains ($3 \mu$m), we assumed above a geometric cross-section for the dust grains (this would not be true in observations of thermal emission at mid-infrared wavelengths; cf. \citealt{Hamilton15}).
The total $m(r)$ is then simply given by the integral of Eq.\:\ref{dm} over $s$.

We now estimate $n(r,s)$ using the results of our semi-analytical investigation of the grains' orbital dynamics from Sec.\:\ref{dynamics}.  
For simplicity, we approximate the shadow of Saturn and its rings as a rectangular prism with cross-section dimensions of $2 R_S \times 2 R_S$.
By using radial bins of equal volume (spaced by $10 R_S$), we ensure that the sought number density $n(r,s)$ is equal to the number of particles we find in each bin to within a normalization constant (which must be fit to the data anyway).  
Since the dimensions of the Phoebe ring are much larger than those of the shadow, our simple choice for the shadow shape does not affect the result.
We can therefore relate $dm(r,s)$ to $N(r,s)$, the number of particles in a Monte Carlo simulation lying in a radial bin centered at $r$, with sizes between $s$ and $s + ds$.  
Writing differentials with $\Delta$ to emphasize the finite size of our bins, we have from Eq.\:\ref{dm},
\begin{equation} \label{4-mphot}
\Delta m(r,s) \: \propto \: s^2 N(r,s) \Delta s,
\end{equation}
where we have assumed that $\varpi_0$ and $P(\alpha)$ are the same for all particles (in our observations, the phase angle $\alpha$ varies from $\approx 3.4^{\circ}$ to $\approx 1.7^{\circ}$ when looking at sections of the Phoebe ring centered at 100 $R_S$ and 200 $R_S$, respectively).  

The strengths of the various relevant perturbations vary with the particle orbits' semimajor axes, which decay according to 
\begin{equation}\label{4-pr}
a = a_0 e^{-t / \tau_{P-R}},
\end{equation}
where $a_0$ is the original semimajor axis, and $\tau_{P-R}$ is the Poynting-Robertson decay timescale \citep{Burns79}.  
Assuming particles share Phoebe's density of 1.6 g/$\text{cm}^3$, 
\begin{equation}\label{taupr}
\tau_{P-R} \approx \Bigg(\frac{s}{7 \mu \text{m}} \Bigg) \text{Myr},
\end{equation}
where $s$ is the particle radius.  
According to Eq. \ref{4-pr}, approximately ln(215/60) $\approx 1.28$ Poynting-Robertson decay timescales are required for particles to approximately reach Iapetus' semimajor axis ($a \approx 60 R_S$) from Phoebe ($a \approx 215 R_S$).  
Because the semimajor axis evolution is the same for all particle sizes if one rescales time through $t' = t / \tau_{P-R}$ (Eq.\:\ref{4-pr}), we chose to consider semimajor axes sampled at one hundred equally spaced $t'$ intervals for all particle sizes:
\begin{equation} \label{deltat}
\Delta t' = \frac{\Delta t}{\tau_{P-R}} = \frac{\text{ln} (215/60)}{100},
\end{equation}
where $\tau_{P-R}$ scales linearly with $s$ (Eq. \ref{taupr}).  

At each of these hundred semimajor axes, we first evaluate the motion of the particle's orbital angular momentum vector in a frame that uses the local Laplace plane as the reference plane (see Sec.\:\ref{4-inc}).  
To a good approximation, the orbital angular momentum vector precesses around the Laplace plane pole at a constant angle given by the free inclination, which is an adiabatic invariant of the motion as the semimajor axis slowly decays \citep{Ward81}.  
At all semimajor axes, we therefore randomly and uniformly sample the orbit's longitude of ascending node on the Laplace plane $\Omega_{Lap}$, and for the free inclination assign Phoebe's current orbital inclination to Saturn's orbital plane of $175.243^{\circ}$ (\url{ssd.jpl.nasa.gov/?sat\_elem}), which very nearly corresponds to the local Laplace plane at Phoebe's orbital radius.
To transform to a common reference frame for all semimajor axes (the frames coinciding with the local Laplace planes are tilted relative to one another as $a$ varies), we calculated the inclination of the local Laplace plane to Saturn's orbit normal at each $a$ \citep{Rosengren14}, and applied the appropriate rotation matrices.

With $a$, $i$ and $\Omega$ (where orbital elements without subscripts are referenced to Saturn's orbital plane) in hand, we proceed to select the eccentricity $e$ and argument of pericenter $\omega$.  
The appropriate level curve that the eccentricity vector follows (Eq.\:\ref{4-H}) is set by the initial conditions.  
Since the escape velocity from Phoebe is small compared to its orbital velocity, and most ejecta is launched at velocities comparable to the escape speed \citep[e.g.,][]{Farinella93}, dust grains will essentially inherit Phoebe's orbital elements at the time of impact.
We therefore set the initial eccentricity to Phoebe's current value (which changes little) and, for computational ease, considered eight equally spaced initial values of $\phi_\odot$ (Sec.\:\ref{4-ecc}).  

As mentioned above, an orbit's angular momentum vector precesses with a constant free inclination about an equilibrium (the local Laplace plane's pole).  
Analogously, (to a good approximation) each orbit's pericenter precesses with a constant free {\it eccentricity} about another equilibrium (the forced eccentricity), i.e., in the polar plots in the top panels of Fig.\:\ref{4-intcomp}, different initial conditions would move on level curves that to first order are concentric circles about the equilibrium forced eccentricity; the radius of the circle is then the constant free eccentricity.  
As in the inclination case with the shifting Laplace plane, the forced eccentricity changes as orbits decay and the relative perturbation strengths vary, and similarly, the free eccentricity is an adiabatic invariant as long as the semimajor-axis decay rate is slow compared to the precession timescale (which is always the case here).

For a given initial condition, we therefore first calculated the approximately conserved free eccentricity.  
Then, at each semimajor axis, we calculated the appropriate forced eccentricity numerically (by finding the point at which level curves collapsed to zero radius), and randomly sampled $e$ and $\phi_\odot$ from a uniform distribution along the perimeter of the level curve.  Then, we obtained $\omega$ using the relationships $\phi_\odot \equiv \varpi - \lambda_\odot$ and $\varpi = \Omega - \omega$.  
Finally, we obtained the last orbital element by selecting the mean anomaly $M$ from a uniform distribution.  

With this procedure, for each of eight equally-spaced values of the initial condition for $\phi_\odot$, and for each of the hundred semimajor axis values, we obtained the orbital elements of particles, and calculated cartesian positions in a system where $z$ points along Saturn's orbit normal, $x$ points from Saturn to the Sun (at the time of observation we are trying to model), and $y$ completes a right-handed triad.  
In order to extract the number of particles along Saturn's shadow, we selected the particles whose positions lay inside the model shadow's rectangular prism, binned by their radial position along $-x$ in slices of length $10 R_S$ from $0-250 R_S$.

The probability of a particle's position falling inside the shadow decreases rapidly with distance from Saturn.  
In order to obtain reliable statistics, we therefore sampled more particles in distant orbits than in tight ones.  
In particular, we calculated the positions of 12500 particles for the innermost semimajor axis at $a=60 R_S$, and boosted the number of sampled particles at each semimajor axis by a factor of $(a / 60 R_S)^3$.  
For a fair comparison, when counting particles in each radial bin, we divided the number of particles from each semimajor axis by the same factor of $(a / 60 R_S)^3$.

Following this procedure, we obtained $N(r_i, s_j, a_k)$, i.e., the number of particles with semimajor axis $a_k$ and size $s_j$ that fell in the bin with radial distance $r_i$, for each of forty different particle sizes\footnote{sampled every micron from 5 to 20 $\mu$m, and at 22, 24, 26, 28, 30, 35, 40, 45, 50, 60, 70, 80, 90, 100, 200, 300, 400, 500, 1000, 2000, 3000, 4000, 5000 and 10000 $\mu$m.}, for each of the 100 sampled semimajor axes.  
The y-scale on our plots is set by the number of particles for which we choose to calculate positions.  
The normalization of our histograms is thus arbitrary, but we obtain an accurate scaling with distance for each particle size.  

The radial distribution of material as a function of particle size, $N(r_i, s_j)$, is then simply given by summing the contributions from each of the semimajor axes; however, knowledge of the relative amounts of material at each semimajor axis requires a model for the injection of particles into the Saturn system (which the data can then support or reject).

We consider here a steady-state model, where Phoebe is bombarded by micrometeoroids at a constant rate, generating $d\dot{N}(s)$ particles with radii between $s$ and $s + ds$ per second\footnote{we note that the implied mass loss rates, even if continued for the age of the solar system, are too small to significantly erode Phoebe.}.  
In our discretized model, within a time $\Delta t$, Phoebe's semimajor axis will receive $N(a=215 R_S,s) = d\dot{N}(s) \times \Delta t$ particles.  
After another $\Delta t$, these particles will have moved to the next semimajor axis in (recall that our semimajor axis values were chosen to each be separated by the same $\Delta t$), and Phoebe's semimajor axis will have received a fresh set of particles.  
After another $\Delta t$, the chain is pushed one link further, until a steady state is reached.  
Thus, each {\it semimajor axis} (i.e., not necessarily each radius) should have the same number of particles.  
We can then simply build the {\it radial} distribution of particles of a given size $N(r_i,s_j)$ by taking the Monte Carlo simulations for grains of radius $s_j$, and for each radial bin adding up equal contributions of particles from each of the hundred sampled semimajor axes, 
\begin{equation}\label{4-sum}
N(r_i,s_j) \: \propto \: \sum_k{N(r_i, s_j, a_k)}.
\end{equation}
Plugging this result into Eq.\:\ref{4-mphot}, we have 
\begin{equation} \label{4-dmrs}
\Delta m(r_i, s_j)\: \propto \:\sum_k{N(r_i, s_j, a_k)s_j^2 \Delta s_j }.
\end{equation}

\subsection{Simulation Results}

\begin{figure}[!ht]
 \centering \resizebox{0.99\columnwidth}{!}{\includegraphics{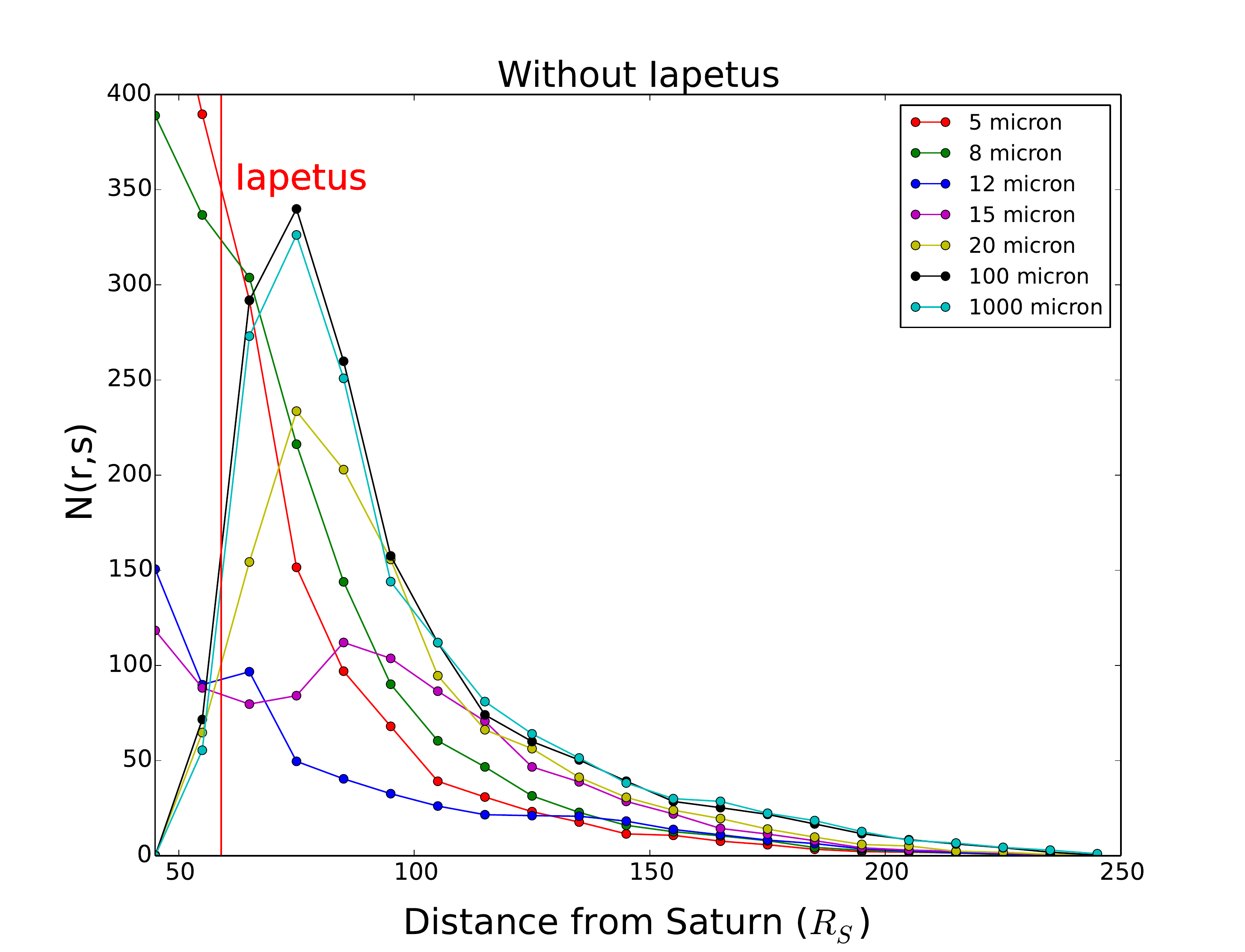}}
\caption{\label{4-pholrad} Number of shadowed particles in the Monte Carlo simulation as a function of orbital radius, for different grain sizes.  
This case does not consider Iapetus sweeping up material (at the vertical red line)---the inner edge around 65 $R_S$ is instead due to the Laplace plane shifting and material tilting off Saturn's orbital plane, so that the planet's shadow does not pierce it.  
As described in the text, particle numbers have been normalized to make up for the fact that we simulated more particles at large distances from Saturn (in order to have a similar chance of finding them in the shadow as grains on tighter orbits).  See the online version for a color figure.}
\end{figure}

Figure \ref{4-pholrad} shows, for different particle sizes, the profile of the number of shadowed particles in the Monte Carlo simulation as a function of radius, $N(r,s)$ (Eq.\:\ref{4-mphot}).  
As one would expect, smaller particles (see the 5 and 8 $\mu$m distributions) reach farther inward, owing to their higher orbital eccentricities induced by radiation pressure.  
By contrast, and of key importance to our later results, large particles ($\gtrsim 20 \mu$m) are relatively unaffected by radiation pressure and converge to a common radial profile.

We note that this case does not consider Iapetus sweeping up material. 
The inner edge around $65 R_S$ for large particles is instead due to our observation's geometry.
At this distance, material tilts off Saturn's orbital plane, so that the shadow no longer passes through the Phoebe ring.  

Intermediate particles (see the 12 and 15 $\mu$m distributions) have peculiar distributions sculpted by complicated dynamics.
As shown by \cite{Rosengren14}, the Laplace surface for particles with these intermediate area-to-mass ratios ``breaks" near the Laplace radius (Eq.\:\ref{rl}), and their orbits are forced to suddenly precess about a more distant equilibrium with a large free inclination.
At such bifurcations, the distributions get substantially puffed up vertically, lowering the number density in the shadow.
This can create local maxima in the distributions of these intermediate-size grains.

\begin{figure}[!ht]
 \centering \resizebox{0.99\columnwidth}{!}{\includegraphics{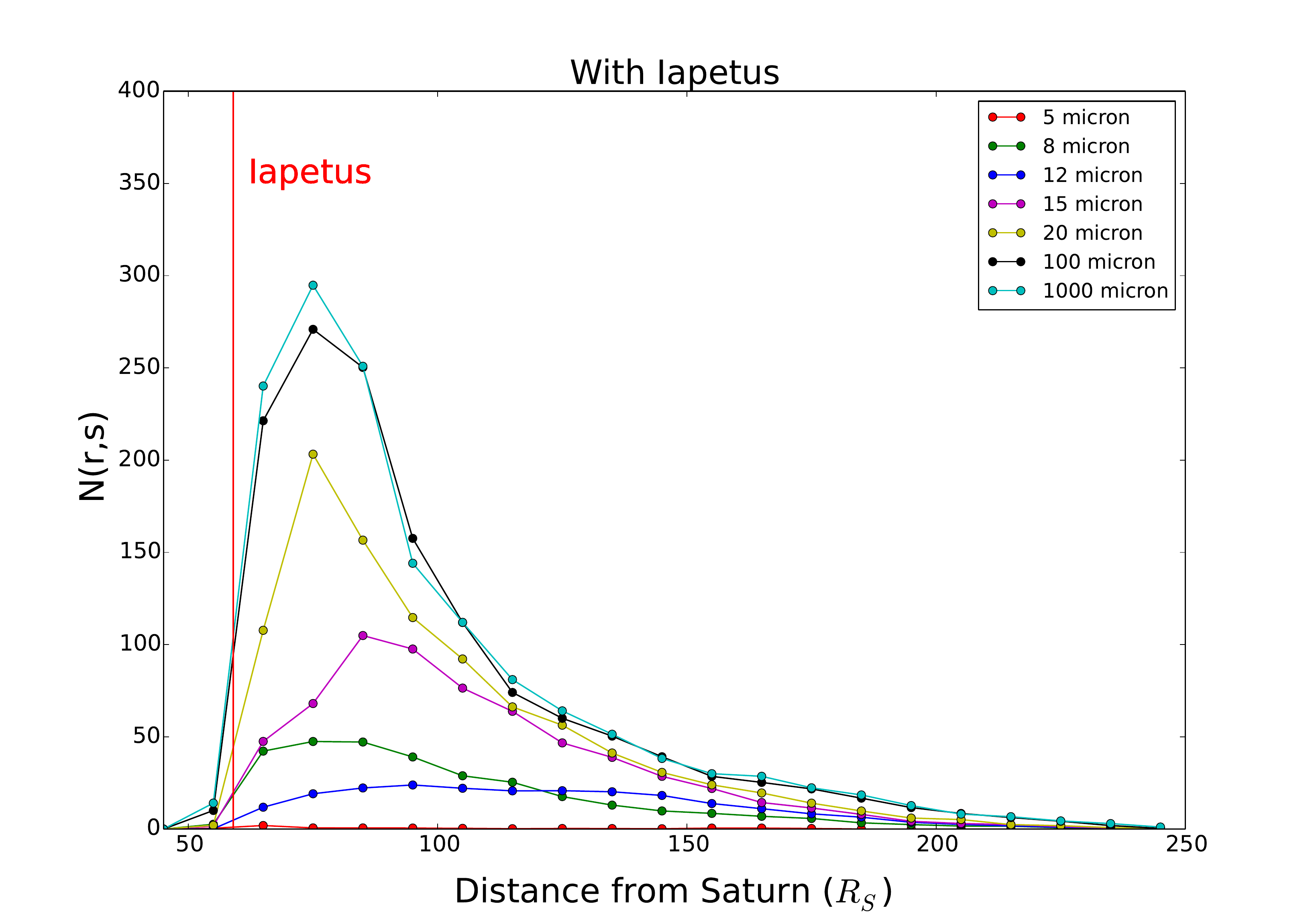}}
\caption{\label{4-pholradIap} Number of shadowed particles in the Monte Carlo simulation as a function of orbital radius, for different grain sizes.  
Any orbits that have pericenters interior to Iapetus' semimajor axis are removed from their respective bins.  See the online version for a color figure.}
\end{figure}

The remaining question is whether Iapetus could cut off the ring at larger radii than $65 R_S$.  
As a limiting case, we imagine that Iapetus sweeps up all the material on orbits with pericenters inside Iapetus' semimajor axis (59 $R_S$).  
This is a good assumption for all but the smallest grains, $s \lesssim 10 \mu m$ \citep{Tamayo11}.  
The result is shown in Fig.\:\ref{4-pholradIap}.

We see that Iapetus only qualitatively affects the radial distributions of small particles ($\lesssim 20 \mu$m).
Starting at approximately $75 R_S$, the orbits of larger grains are tilted off the orbital plane that is pierced by Saturn's shadow, so they disappear from our observations before we can observe Iapetus sweeping them up.
By contrast, the more complicated dynamics of small grains ($\lesssim 20 \mu$m) causes some material to remain in the shadow closer to Saturn, making it possible for our observational setup to see the effect of Iapetus intercepting these diminutive particles.
Additionally, small grains develop large orbital eccentricities through radiation pressure, and are therefore able to reach Iapetus at pericenter from larger Saturnocentric distances.
Of course, in reality, Iapetus will sweep up material of all sizes; in fact, larger particles are more likely to be intercepted since they decay inward more slowly through Poynting-Robertson drag \citep{Tamayo11}.
We now compare these distributions to the observations.


\section{Comparing Theoretical Models to the Data} \label{comparison}
To connect our theoretical radial distributions for various particle sizes (Figs.\:\ref{4-pholrad} and \ref{4-pholradIap}) with the observed photometry, one must combine the $\Delta m(r_i,s_j)$ into a single $m(r_i)$ (Eq.\:\ref{4-dmrs}).  
In addition to any intrinsic particle size distribution, because smaller particles evolve inward faster than large grains (Eq.\:\ref{deltat}), we must consider that a given semimajor axis will receive more small grains than large ones in a given time interval.  
To this end, we take the input {\it rate} of particles per unit time at Phoebe's semimajor axis (which is the same for all our $a$ values in a steady state) to follow a power-law distribution with index $-q$, 
\begin{equation} \label{sizedist}
\dot{N}(s_j, a_k) \:\propto\: s_j^{-q} \Delta s_j.
\end{equation}
Then, since each of our hundred sampled semimajor axes are separated by the same (size-dependent) $\Delta t$ (Eq.\:\ref{deltat}), we can obtain the number of particles of size $s_j$ of semimajor axis $a_k$ $N(s_j, a_k)$ in our discretized model through
\begin{equation} \label{4-w}
N(s_j, a_k) \: \propto \: \Delta t \times s_j^{-q} \Delta s_j \:\propto\: s_j^{-(q-1)} \Delta s_j,
\end{equation}
where the additional factor of $s$ comes from the factor of $\tau_{P-R}$ (Eq.\:\ref{4-pr}) in $\Delta t$ from Eq.\:\ref{deltat}.  
Therefore, in combining particle sizes (assuming a steady state), one should weight the contribution from each grain radius by a factor $w_j = s_j^{-(q-1)} \Delta s_j$.  Since $w_j$ is independent of $a$, we can obtain $m(r_i)$ directly from Eq.\:\ref{4-dmrs},
\begin{equation} \label{4-finalm}
m(r_i) \:\propto\: \sum_{j = s_{min}}^{s_{max}} \Delta m(r_i, s_j) w_j \:\propto\: \sum_{j = s_{min}}^{s_{max}} \sum_k N(r_i, s_j, a_k)s^{3-q} \Delta s_j.
\end{equation}
In summary, Eq.\:\ref{4-finalm} relates the number of particles in each bin of our Monte Carlo simulations to the $m(r_i)$ that we use to convert our modeled pathlengths through each radial slice of the shadow into the expected brightness deficit in a particular pixel (see Fig.\:\ref{4-sub}).

\begin{figure}[!ht]
 \centering \resizebox{0.99\columnwidth}{!}{\includegraphics{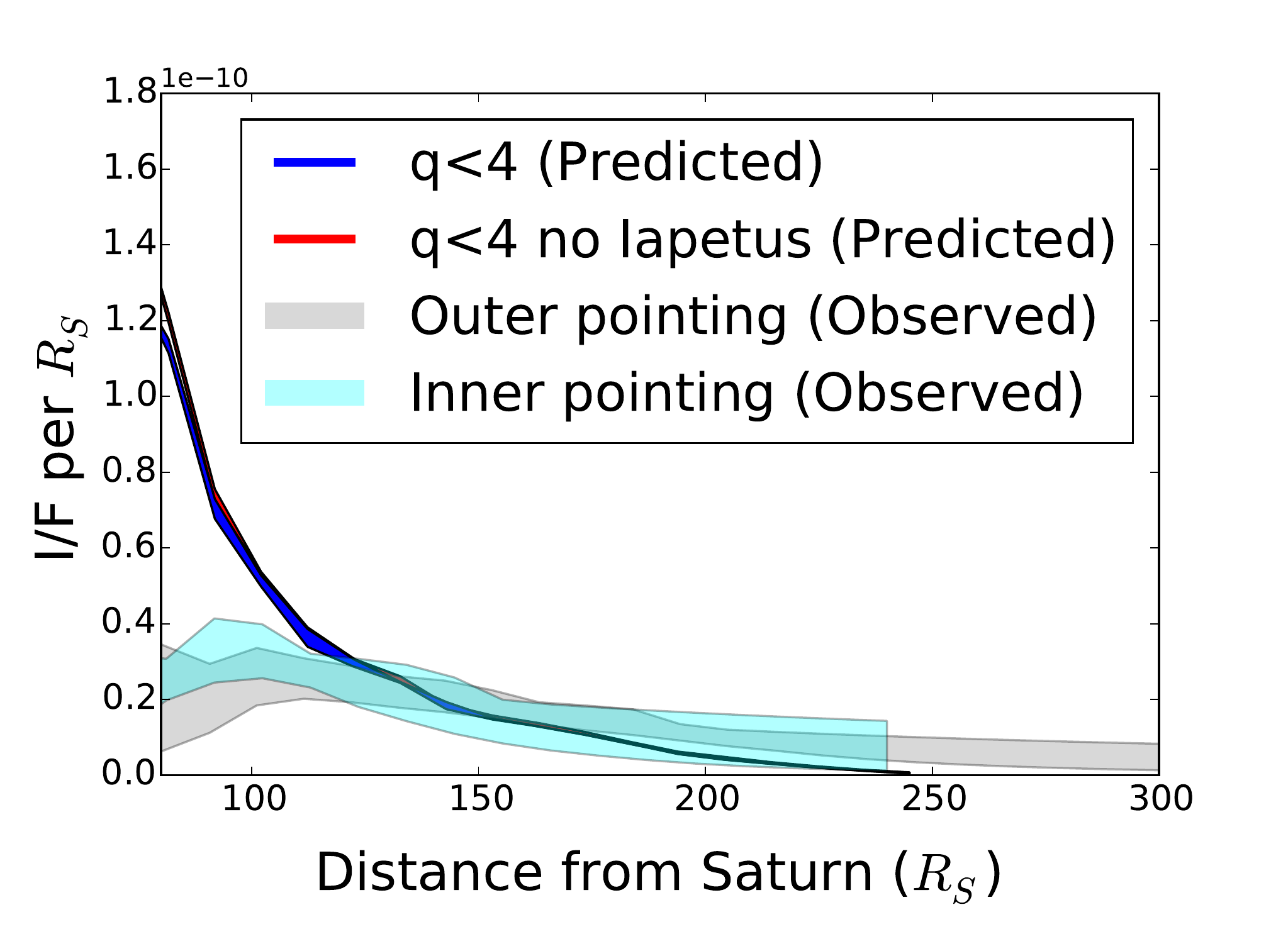}}
\caption{\label{powerlaws} Predicted I/F per $R_S$ as a function of radial distance from Saturn.  
The red line shows the predicted radial profile without accounting for Iapetus sweeping up material.
The blue line removes any material whose orbit crosses Iapetus (at $59R_S$).  
In gray and cyan are the observed distributions plotted in the right panel of Fig.\:\ref{best-vertline}.
The predicted profiles have been normalized to agree with the observations at $140 R_S$.}
\end{figure}

Figure \:\ref{powerlaws} shows our theoretical radial profiles for typical size distributions of solar system rings (power law indices $\leq 4$).
In particular, we fill the space between the curves for a distribution with $q=1$ and $q=4$, using a minimum and maximum particle size of 5$\mu$m and 1 cm, respectively.  
The agreement for power laws $\leq 4$ is due to the factor of $s^{3-q}$ in Eq.\:\ref{4-finalm}, which highlights large particles that all have similar dynamics (only for indices steeper than $q=4$ would small particles with different dynamical behaviors begin to dominate).

Additionally, we see that, for these size distributions, our models are insensitive to Iapetus sweeping up material.
This is because for $q\leq4$, the dominant large grains are too large to be significantly affected by radiation pressure, and therefore do not have sufficient orbital eccentricity to reach Iapetus (at $59 R_S$) from the larger distances spanned by our observations.

While our observations do not reach inward to the distance from Saturn where large particles are removed from the shadow ($\approx 75 R_S$, see Fig.\:\ref{4-pholrad}), our observed radial profiles (gray and cyan) seem to plateau and possibly dip starting at $100 R_S$.

\subsection{Phoebe ring grains are small}

Taking a step back, we first note that the blue and red curves in Fig.\:\ref{powerlaws} simply follow the radial distribution to which dust grains of increasing size converge as radiation pressure plays a decreasingly important role (compare with the 20, 100 and 1000 $\mu$m curves in Fig.\:\ref{4-pholrad}, in the range beyond 80 $R_S$ from Saturn).
In this regime of large grains, radiation pressure represents a small perturbation.
This renders the approximations in our semi-analytic model (Sec.\:\ref{dynamics}) excellent, which should yield accurate predicted radial profiles.
The fact that the observed radial profile instead plateaus at $\sim 100 R_S$ therefore implies that particles $\gtrsim 20\mu$m cannot dominate the scattered light flux from the Phoebe ring\footnote{\cite{Hamilton15} recently reached a similar conclusion from infrared observations with WISE.}.

It is less clear {\it why} small grains dominate the scattered light flux.
If we assume as above that Phoebe ring grains are produced in a steady state, this suggests a steep power law size distribution with index $q>4$, which would be unusual among rings in the Solar System \citep{Burns01}.
However, the Phoebe ring has a substantially lower normal optical depth than known planetary rings.
This implies that dust grains should not collide with one another, even over the $\sim$ Myr timescales required for material to decay inward through Poynting-Robertson drag.  
The steep inferred particle size distributions could therefore reflect the initial size distribution of ejecta.
One way to observationally test whether the Phoebe ring indeed has such an unusually steep size distribution would be to measure its brightness in different optical filters; such a ring should appear blue.  

Alternatively, release of all the Phoebe ring material in a single large collision would likely admit shallower size distributions (see the preceding section), though one would have to additionally model and fit for the time of the event.
Another possibility is that other processes could be preferentially destroying large grains.
The P-R decay timescales are much longer (at least 1 Myr, see Eq.\:\ref{taupr}) than the lifetimes of dust particles in typical planetary rings deep in the host's magnetosphere \citep{Burns01}, so one might expect different effects to dominate in this unusual regime.
In particular, micrometeoroids should preferentially break up larger grains \citep{Burns01}.
Finally, the solar wind could be affecting grains as they evolve inward outside Saturn's magnetosphere.

The above conclusions suggest one should fit the Phoebe ring's radial profile with small grains / steep power laws.  
The problem is that our approximations from Sec.\:\ref{dynamics} are much poorer for these diminutive grains.
Not only do these particles acquire substantial orbital eccentricities, the Laplace equilibria for intermediate particles $\approx 10-15 \mu$m become unstable as they evolve inward, leaving orbits with large free inclinations to their newfound centers \citep{Rosengren14}. 

\begin{figure}[!ht]
 \centering \resizebox{0.99\columnwidth}{!}{\includegraphics{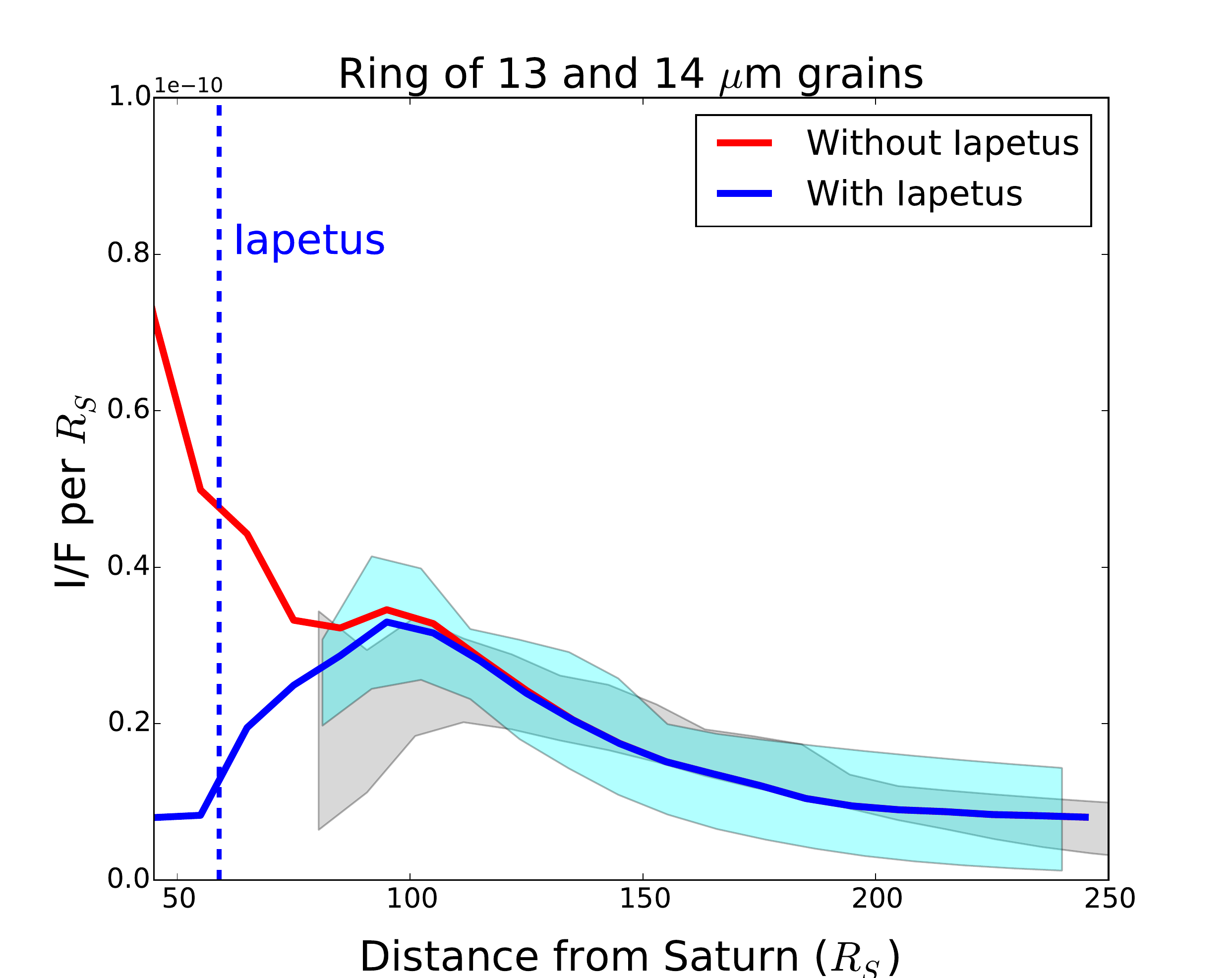}}
\caption{\label{plateau} Number of shadowed particles in the Monte Carlo simulation, assuming an equal number of 13 and 14 $\mu$m grains, and a constant I/F offset of $8\times 10^{-12}$ from other irregular satellites.}
\end{figure}

This presents another promising extension of this work, since we see hints in our models that such phenomena could explain the observed ``plateau."
For example, as shown in Fig.\:\ref{plateau}, a ring composed entirely of 13 and 14$\mu$m grains gives a reasonable qualitative match to the data.
One can also see that Iapetus strongly sculpts the distribution inside $\approx 100 R_S$.  
Given the remarkably rich dynamics, one would have to carry out a suite of numerical integrations to accurately compare this model to the observations. 
Additionally, to separate these complicated dynamical effects from Iapetus sweeping up material, one would likely have to accurately model collisions with the satellite.
We defer this numerical effort to future work, but note the importance of pushing observations inward to $\sim 75 R_S$.  

\subsection{Contributions to the Phoebe ring from other irregular satellites}

As can also be seen in Fig.\:\ref{powerlaws}, our theoretical models predict less scattered light at Phoebe's apocenter and beyond ($> 250 R_S$) than shown by the data.
The physical reason is that for retrograde orbits, radiation pressure induces a forced eccentricity (Sec.\:\ref{4-ecc}) that is directed {\it away} from the Sun along $\phi_\odot = \pi$ (see Fig.\:\ref{4-intcomp}).
This means that when particle orbits reach the maximum eccentricity in their secular cycle, their pericenters point away from the Sun, i.e. along the shadow axis.
Our observations therefore always sample the most eccentric particles at pericenter---this skews the radial distributions toward smaller distances from Saturn, leaving little material beyond Phoebe's apocenter.\footnote{Were Phoebe ring grains to instead orbit in a prograde direction, the behavior would be opposite and particles would fill the shadow to much larger distances \citep{Hamilton96Mars}.}

As \cite{Hamilton15} also argue from infrared WISE data, this suggests that other irregular satellites also contribute to the ``Phoebe" ring.  
The more distant irregular satellites Ymir, Suttungr, Thrymr and Greip are promising candidates, given their similar orbital inclinations to Phoebe (since this determines the derived vertical disk thickness).
There are likely also additional bodies too small to detect with current technology.
The contributions from other irregular satellites merits further study, since our empirical model fit a single power law at large distances.



\section{Conclusion} \label{conc}
By measuring the deficit in scattered light from Phoebe ring grains in Saturn's shadow, we were able to reconstruct a radial profile of material in this vast debris disk (Fig.\:\ref{best-vertline}).
We also obtained an integrated I/F at 0.635 $\mu$m along Saturn's shadow from 80-250 $R_S$ of $2.7^{+0.9}_{-0.3} \times 10^{-9}$.
To date, only this technique has yielded measurements of the Phoebe ring at optical wavelengths.
Additionally, the method's inherent attenuation of scattered light from Saturn makes it possible to probe material closer to Saturn than has been feasible with infrared observatories in orbit around Earth.

Combining such a measurement of scattered light at optical wavelengths with ones of thermal emission at infrared wavelengths, like those of \cite{Verbiscer09} and \cite{Hamilton15}, allow one to estimate the particle albedos.  
This was done by THB14, who found grain albedos consistent with dark ejecta from Phoebe.
However, while the data presented in this paper are substantially better than those analyzed by THB14, there remain large uncertainties in the particles' infrared emissivities and phase functions  (though see \citealt{Hedman15} for recent progress).
We therefore defer an improved analysis until more observations at undertaken at new wavelengths.

We find that the scattered light signal rises as one moves inward from Phoebe to Saturn (as expected), but then ``plateaus" at $\approx 100 R_S$ (Fig.\:\ref{plateau}).  
We developed a semi-analytic treatment for the size-dependent dust dynamics of Phoebe ring grains, and used this to generate a Monte Carlo model of the material in Phoebe ring's shadow.
Our models, which should be accurate for grains $\gtrsim 20\mu$m in size, deviate from the ring's observed radial profile inside $\approx 100 R_S$.
We conclude that the Phoebe ring's scattered light signal must be dominated by small dust grains ($\lesssim 20\mu$m).
Assuming the Phoebe ring is generated through a steady-state process of micrometeoroid bombardment, this implies that a particle size distribution with an index $> 4$, which is unusually steep among solar system rings.
This agrees with a recent analysis of the ring's infrared thermal emission with WISE \citep{Hamilton15}.  
Again in agreement with \cite{Hamilton15}, we find that additional irregular satellites beyond Phoebe must contribute material to the ``Phoebe ring," in order to explain the observed fluxes at and beyond Phoebe's apocenter.

The lack of large particles in the Phoebe ring may have important implications.
Because the optical depth is so low, particles smaller than $\sim 100 \mu$m should not suffer mutual collisions \citep{Tamayo11}.
The steep size distribution may therefore trace the original size distribution of ejecta from micrometeoroid bombardment.
Alternatively, the small particles may suggest that another process, perhaps micrometeorite bombardment, preferentially breaks apart large grains as they more slowly decay inward over several Myr.  

It is unclear whether the ``plateau" feature we observe is due to Iapetus sweeping up material, the complicated dynamics of small dust grains, or both.
The approximations in our analytical model break down for these diminutive particles, so an in-depth numerical study will be required to accurately untangle these effects.
Nevertheless, it is theoretically expected that Iapetus should efficiently sweep up particles $\gtrsim 10 \mu$m \citep{Tamayo11}, so it will be a valuable task to push future data analysis and modeling efforts to observationally test how well Iapetus carves out the inner edge to the Phoebe ring.

\section{Acknowledgments}
We would like to thank Michael W. Evans, Philip D. Nicholson and Matthew S. Tiscareno for technical help and insightful discussions.  This research was partially supported by a postdoctoral fellowship from the Centre for Planetary Sciences at the University of Toronto at Scarborough, and we gratefully acknowledge support from the Cassini mission.

\appendix
\section{Testing the pipeline}

We have tested our procedures by running synthetic images through our pipeline.  
Using our modeled pathlengths for the outer pointing in Rev 197, and assuming the broken power-law model for the Phoebe ring of Sec.~\ref{brokenpowerlaw}, we constructed images with the expected dimming for shadowed pixels.  
Superimposing Gaussian noise with mean I/F $5\times 10^{-8}$, and standard deviation $10^{-8}$ (the values we found in the same data set after filtering bad pixels), we ran these fake images through the pipeline to try and retrieve the input radial model.

\begin{figure}[!ht]
 \centering \resizebox{0.99\columnwidth}{!}{\includegraphics{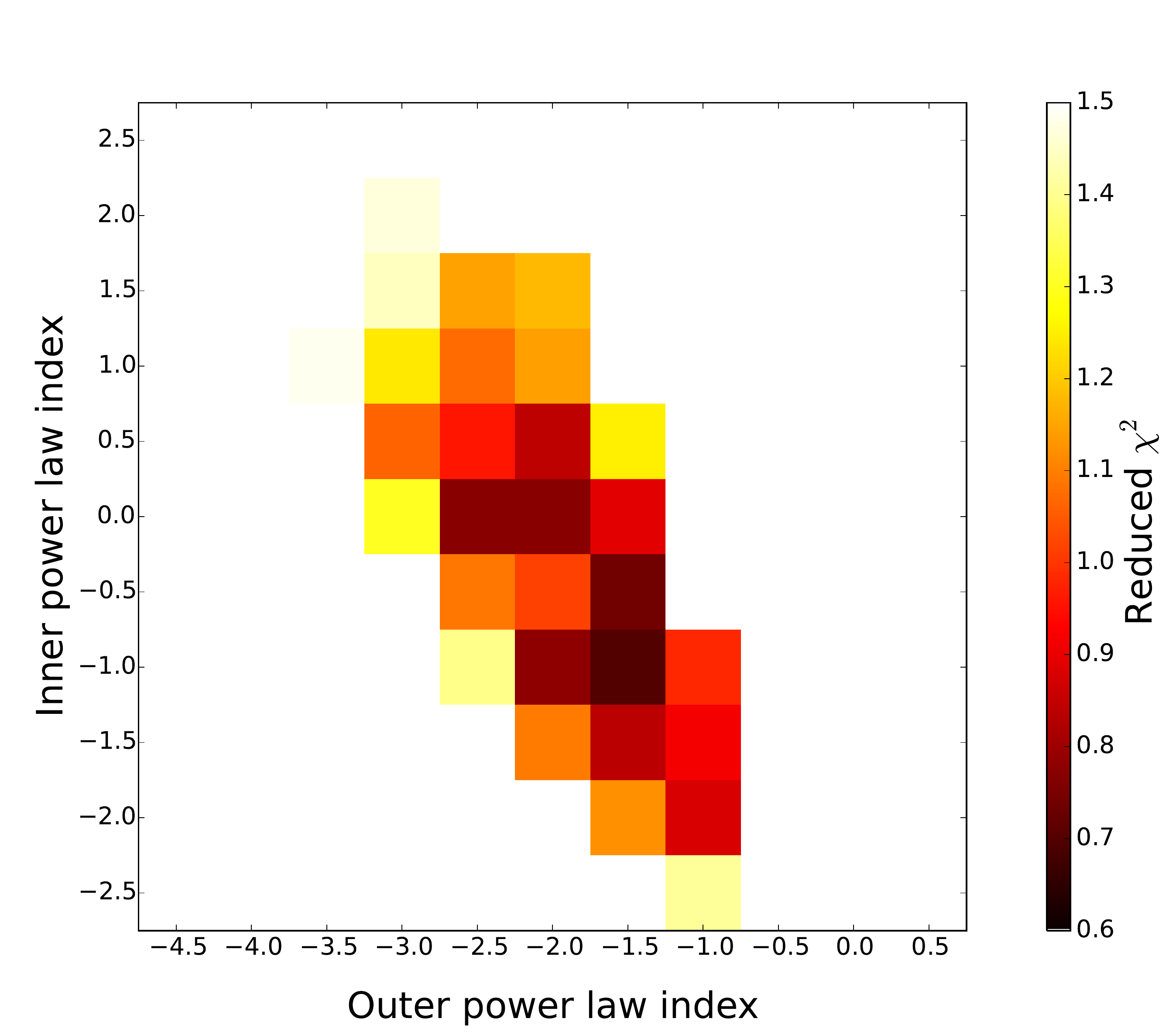}}
\caption{\label{testminchi}  Distribution of reduced $\chi^2$ values for a grid of models, applied to a synthetic dataset overlaid with Gaussian noise that was generated using the model parameters at the center of the grid.  
See text for discussion.}
\end{figure}

In particular, we generated synthetic images from a model with $n_{\text{Inner}} = 0$, $n_{\text{Outer}} = -2$, and $R_k = 130 R_S$ (Eq.~\ref{brokenplaweqs}).  
We then calculated reduced $\chi^2$ values using our pipeline for a grid of power law indices centered on the input values (with the break fixed at 130 $R_S$).  
The result is shown in Fig.\:\ref{testminchi}.

Several features stand out in this plot.  
Perhaps most striking is that the input model does not yield the lowest $\chi^2$.  Additionally, the $\chi^2$ is systematically below unity for the best models, and does not vary smoothly across the grid.  
We address these points in reverse order.

The reduced $\chi^2$ does not vary smoothly primarily because of our selection of the best-fit normalization.  
In reality, this slice for a break location at $130 R_S$ is three-dimensional (specifying the normalization and two power-law indices).  
If one pictures Fig.\:\ref{testminchi} as extending into the page along the normalization direction, the plotted colors correspond to the minimum chi squared value from the ``column" of normalizations ``below" each grid point.  
Because in general the minimum $\chi^2$ values lie at different ``depths," the variation is not smooth across the grid.  
A similar effect can be seen in Fig. 9 of \cite{Nicholson14}.  

We attribute the systematically low reduced $\chi^2$ values to an overestimation of our number of degrees of freedom, which can often be a problem for non-linear models \citep[e.g.,][]{Andrae10}.  
Due to the large number of pixels in each of our datasets ($\sim 10^7$), we choose to do a simple and thus necessarily rough statistical analysis.  
We generated 300 fake images using the same broken power-law model given above, each with a different Gaussian noise realization (with parameters as above).  
We then ran each synthetic data set through our pipeline, fitting to the same input model used to generate the fake images.
To additionally test our procedure's ability to extract the correct normalization, we initially guess a normalization that is two times too large.

Our procedure systematically retrieves the correct normalization to within 0.06\% (mean), with a standard deviation of 0.3\%.  
A na{\"i}ve counting of the number of bins entering our $\chi^2$ evaluation suggests 259 degrees of freedom.  
We find that a histogram of our $\chi^2$ values is instead best fit by a $\chi^2$ distribution with 209 degres of freedom.  
A Komolgorov-Smirnov test gives a $p$ value of 0.43 that our histogram is drawn from such a $\chi^2$ distribution (with $\approx 80\%$ the number of degrees of freedom one would na{\"i}vely estimate).  
While we might thus adjust the degrees of freedom in our analysis by $80\%$ to evaluate the probabilities entering our marginalizations, we nevertheless choose to normalize all reduced $\chi^2$ values to the minimum value.  
We find that if we raise the reduced $\chi^2$ values by $20\%$, our marginalized estimates are extremely sensitive to the few parameter combinations lying at the bottom of the deepest valleys of $\chi^2$ space.
Normalizing to the minimum value more equitably samples the best fits to the data and seems like a more balanced representation of the models, given that our underlying statistical analysis is approximate.
We partially compensate for this by making conservative estimates of the errors on the parameters we extract, bracketing the wide range in parameter space that yields reasonable fits.

Finally, we consider that the input model does not yield the lowest $\chi^2$.  
This reflects the fact that in order to obtain enough signal, our geometry is such that we look nearly radially outward down the axis of the shadow.  
This fundamentally limits the amount of radial information that we can extract from our data.
Thus, the models along the band of low $\chi^2$ values are all good models that approximately conserve the integrated amount of material in a column along the line of sight.
Superposed on this band of good models is the statistical variation one would expect for a $\chi^2$ distribution, which, as argued above, is jumpy because we are probing to different ``depths" along the normalization direction in parameter space.  
It is thus not surprising that one of the equally good models near the input model would statistically have a lower $\chi^2$. 
If we try the same analysis with a different noise realization, we find the same band of low $\chi^2$ values, with the same dispersion, but a different grid point along the band becomes the ``best fit."

As a final consideration, we investigate whether the low $\chi^2$ values might imply we are overfitting the data.
To test this, we perform our normal procedure, but only on the even images, obtaining a best-fit model (including a value for the normalization).  
We then calculate a reduced $\chi^2$ value for that model (this time with the normalization fixed), using the odd images.
If we were fitting noise in the even images, the $\chi^2$ would suffer in the odd images, but we find that the reduced $\chi^2$ values are statistically indistinguishable between the even and odd images, and both look like Fig.~\ref{testminchi}.

\bibliography{/Users/dtamayo/Documents/Research/papers/Bib}

\end{document}